\documentclass[
 rsi,
 amsmath,amssymb,
 preprint,
 reprint,
 author-numerical,
]{revtex4-2}

\usepackage[utf8]{inputenc}
\usepackage[T1]{fontenc}
\usepackage{mathptmx}
\usepackage{etoolbox}
\usepackage{dcolumn}
\usepackage{bm}
\usepackage{orcidlink}
\usepackage{graphicx,subcaption}
\usepackage{siunitx}
\usepackage{braket}
\usepackage{amssymb}
\usepackage{amsmath}
\usepackage{sidecap}
\usepackage{changes}
\usepackage{todonotes}
\usepackage{caption}
\usepackage{subcaption}
\setuptodonotes{inline, color=red!30}
\usepackage[version=4]{mhchem}
\usepackage{nicefrac}
\usepackage{bbold}
\usepackage{placeins}
\usepackage{xfrac}
\usepackage[scr=boondox]{mathalpha}
\usepackage{babel}

\newcommand{\Rb}{\ce{^{87}Rb}}
\newcommand{\g}{\ket{\uparrow}}
\newcommand{\gaux}{\ket{\downarrow}}
\newcommand{\gs}{\ket{\mathrm{g}}}
\newcommand{\inter}{\ket{\mathrm{e}}}
\newcommand{\intp}{\ket{\mathrm{+}}}
\newcommand{\intn}{\ket{\mathrm{-}}}
\newcommand{\ryd}{\ket{\mathrm{r}}}
\newcommand{\rydaux}{\ket{\mathrm{r}^*}}
\newcommand{\rydb}{\ket{\mathrm{r}^{(*)}}}
\newcommand{\op}{\Omega_p}
\newcommand{\Oc}{\Omega_c}
\newcommand{\Dp}{\Delta_p}
\newcommand{\Dc}{\delta_c}
\newcommand{\vint}{V_\mathrm{RyRy}}
\newcommand{\ve}{V_\mathrm{int}}
\newcommand{\scattot}{R_\mathrm{sc}}
\newcommand{\Hryd}{\hat{H}_\mathrm{RyRy}}
\newcommand{\Hrabi}[1][]{\hat{H}_\mathrm{Rabi #1}}
\newcommand{\Hhfs}[1][]{\hat{H}_\mathrm{HFS #1}}

\newcommand{\Hati}{\hat{H}_\mathrm{ATI}}
\newcommand{\base}[1][]{\ket{\Psi_{#1}}}
\newcommand{\psiq}[1][]{\ket{\psi_{#1}}_\mathrm{C}}
\newcommand{\psis}[1][]{\ket{\psi_{#1}}_\mathrm{S}}
\newcommand{\psig}[1][]{\ket{\psi_{#1}}_\mathrm{G}}
\newcommand{\psiqfs}{\ket{n, l, j, m_j}_\mathrm{C}}
\newcommand{\psisfs}{\ket{n, l, j, m_j}_\mathrm{S}}
\newcommand{\psisfss}{\ket{n, l, j, m_j, I, m_I}_\mathrm{S}}
\newcommand{\ident}[1][]{\mathbb{1}_{#1}}
\newcommand{\sep}{R}
\newcommand{\egbasefs}[1][]{\ket{\tilde{\phi}_{#1}}}
\newcommand{\egbase}[1][]{\ket{\phi_{#1}}}
\newcommand{\egbasebra}[1][]{\bra{\phi_{#1}}}
\newcommand{\dmatrix}[1][]{\hat{\rho}_{#1}}
\newcommand{\dmes}[1][]{\ket{\Theta_{#1}}}
\newcommand{\nscat}{N^{\left(g\right)}}
\newcommand{\tp}{t_p}
\newcommand{\branch}[1][]{B_{#1}}
\newcommand{\transferr}{P_\mathrm{tr}}
\newcommand{\transfer}{\mathcal{P}}
\newcommand{\ratio}{\mathcal{R}}
\newcommand{\dens}{\rho_{\sep}}

\newcommand{\gsnum}[1][]{\ket{5\text{S}_{1/2};\,F=2 #1}}
\newcommand{\gsnumaux}[1][]{\ket{5\text{S}_{1/2};\,F=1 #1}}
\newcommand{\intnum}[1][]{\ket{6\text{P}_{3/2} #1}}
\newcommand{\rydnumaux}[1][]{\ket{38\text{S}_{1/2} #1}}
\newcommand{\rydnum}[1][]{\ket{39\text{S}_{1/2} #1}}
\newcommand{\rydfo}[1][]{\ket{38\text{P}_{3/2} #1}}
\newcommand{\dipoleop}{\mathbf{d}\kern-0.425em\hat{\phantom{\mathbf{d}}}}
\newcommand{\pois}{\mathrm{Pois}}
\newcommand{\Nph}{N_\mathrm{ph}}
\newcommand{\Nato}{N_\mathrm{at}}
\newcommand{\nph}{n_\mathrm{ph}}
\newcommand{\nato}{n_\mathrm{at}}

\begin{document}

\title{Fast and robust detection of single Rydberg excitations in mesoscopic ensembles}
\author{Sven Schmidt~\orcidlink{0009-0000-5626-2630}}
\affiliation{Department of Physics and Research Center OPTIMAS, Rheinland-Pfälzische Technische Universität Kaiserslautern-Landau, 67663 Kaiserslautern, Germany}
\email{agott-publication@physik.rptu.de}

\author{Aaron Thielmann~\orcidlink{0009-0000-0799-9927}}
\affiliation{Department of Physics and Research Center OPTIMAS, Rheinland-Pfälzische Technische Universität Kaiserslautern-Landau, 67663 Kaiserslautern, Germany}

\author{Thomas Niederprüm~\orcidlink{0000-0001-8336-4667}}

\author{Herwig Ott~\orcidlink{0000-0002-3155-2719}}
\affiliation{Department of Physics and Research Center OPTIMAS, Rheinland-Pfälzische Technische Universität Kaiserslautern-Landau, 67663 Kaiserslautern, Germany}

\date{\today}

\begin{abstract}
We propose a novel non-destructive method for the detection of single Rydberg excitations in a mesoscopic ensemble. The protocol achieves high fidelities on a microsecond timescale and is robust against changes in the probe laser frequency.
The technique relies on optical pumping in Autler Townes configuration, whose efficiency is controlled by the presence/absence of a Rydberg excitation. 
Taking rubidium atoms as an example, we give realistic estimates for the achievable fidelities and parameters. However, our protocol can be transferred to any other atomic species which features multiple stable states.  
Our protocol is applicable in quantum simulation and quantum information processing with mesoscopic ensembles requiring fast and high fidelity Rydberg state detection.  
\end{abstract}

\maketitle

    \section{Introduction}
    \label{sec:intro}

    Due to their outstanding properties in terms of scalability and controllability, tweezer arrays of neutral atoms emerged as one of the prime candidates for quantum simulation and quantum information processing in past years \cite{doi:10.1126/science.aah3778,PhysRevX.8.041054,PhysRevX.8.041055}.
    The building principle is based on the generation of arbitrary configurations of optical tweezers with single site resolution. 
    Although separated by several micrometers, excitations to high lying Rydberg states can be used to mediate tunable interactions between the traps \cite{PhysRevLett.85.2208, RevModPhys.82.2313}, making use of both, well isolated sites and versatile couplings.
    Based on this concept, groundbreaking progress such as the simulation of quantum spin models with hundreds of atoms \cite{Ebadi2021-li, Scholl2021}, realization of high fidelity quantum gates \cite{PhysRevLett.123.170503, PhysRevLett.123.230501, PhysRevLett.121.123603, Evered2023} and implementation of quantum algorithms \cite{Graham2022, Bluvstein2022} has already been demonstrated, with recent findings underlining its scalability by achieving arrays of a few thousand sites \cite{Pause:24, manetsch2024tweezerarray6100highly}.
    Although most commonly loaded with a single atom per site, ensembles with a few (hundred) atoms \cite{ Wang2020, PhysRevA.106.022604} in each tweezer can exploit collective excitations \cite{PhysRevLett.115.093601, PhysRevLett.127.063604}, speeding up the preparation and detection \cite{PhysRevLett.127.050501}.
    While several suggestions for fast detection in the $\SI{}{\micro\second}$ timescale have already been made, they either lack a sufficient spatial resolution \cite{PhysRevA.97.063613}, or require large ensembles of hundreds of atoms and specialized detection setups with near single photon sensitivity \cite{PhysRevLett.127.050501}.
    We aim to bridge the gap between the well established methods of fluorescence imaging of single atoms on the $\SI{}{\milli\second}$ scale \cite{Madjarov2020, PhysRevLett.133.013401} and those fast techniques by amplifying the signal of a single excitation by a surrounding ensemble, reducing the state read out time and, optionally, storing its information in the ensemble.

    Our method is designed for small atomic ensembles and employs a two-photon transition from the ground to the Rydberg state in Autler--Townes configuration.
    In the presence of a Rydberg excitation in the sample, optical pumping of the sample into a second hyperfine ground state is activated.
    This leads to a very fast projective non-destructive measurement of the Rydberg excitation, which is mapped to the ground state population on a $\SI{}{\micro\second}$ scale.
    The population of the ground state is then read out immediately afterwards or later on.
    This protocol is fast, provides high imaging fidelity and even allows for the detection of multiple subsequent Rydberg excitations in the same sample.
    
     \section{Basic principle}
    \label{sec:basic_principle}
    
\begin{figure}[ht]
        \captionsetup[subfigure]{oneside,margin={0.5cm,0cm}}
        \vspace*{10pt}
        \centering
        \begin{subfigure}[t]{0.225\textwidth}
            \caption{}
            \centering
            \includegraphics[width=0.5\textwidth]{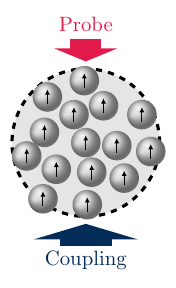}
            \label{sub_fig:idea_sketch_off}
        \end{subfigure}
        \begin{subfigure}[t]{0.225\textwidth}
            \caption{}
            \centering
            \includegraphics[width=0.5\textwidth]{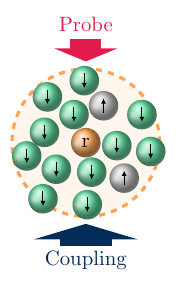}
            \label{sub_fig:idea_sketch_on}
        \end{subfigure}
        \vspace*{-10pt}
        \caption{The ensemble prepared in $\g$ (gray) is coupled to an auxiliary Rydberg state by a two-photon transition, combining an intense coupling (blue) and weak probe (red) light in Autler Townes configuration. 
        (a) Without a Rydberg excitation present, strong interaction with the coupling light screens the atoms from the probe, retaining their initial state. 
        (b) In presence of a Rydberg excitation $\ryd$ (orange), the Autler Townes condition is broken and scattering on the probe transition takes place. Atoms are transferred to $\gaux$ (green).}
        \label{fig:idea_sketch}
    \end{figure}

    For the imaging protocol to work, we require an atomic species with at least two (meta-)stable states while we limit the following discussion to hyperfine ground states. The sample is initially prepared in one of the ground states $\g$ (Fig.\,\ref{sub_fig:idea_sketch_off}).
    In order to detect a Rydberg state $\ryd$ in the atomic sample, we couple the ground state to an auxiliary Rydberg state $\rydaux$ via an intermediate state $\inter$ in a two-photon transition.
    The coupling is realized in Autler-Townes configuration, i.e. the coupling on the upper transition to the Rydberg state is much stronger than the radiative decay rate of the intermediate state.
    When both light fields are resonant, the Autler-Townes splitting of the intermediate state strongly suppresses scattering of the probe laser~\cite{PhysRev.100.703}.
    Consequently, optical pumping to a second hyperfine ground state $\gaux$ is strongly suppressed.
    In the presence of a Rydberg excitation $\ryd$ in the sample (control atom) the energy of the auxiliary Rydberg state $\rydaux$ is shifted due to the dipole--dipole interaction and the Autler--Townes condition breaks down.  
    Optical pumping is now active and ground state population is quickly transferred to the state $\gaux$.

\begin{figure}[ht]
        \captionsetup[subfigure]{oneside,margin={0.5cm,0cm}}
        \vspace*{10pt}
        \centering
        \begin{subfigure}[t]{0.225\textwidth}
            \caption{}
            \centering
            \includegraphics[width=\textwidth]{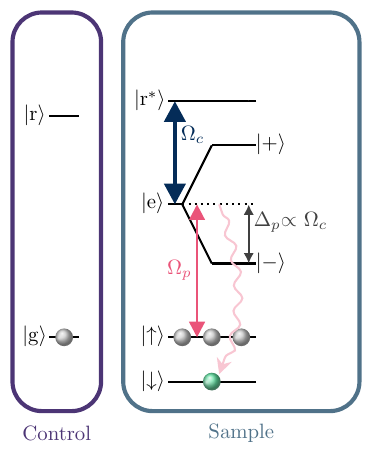}
            \label{sub_fig:scheme_off}
        \end{subfigure}
        \begin{subfigure}[t]{0.225\textwidth}
            \caption{}
            \centering
            \includegraphics[width=\textwidth]{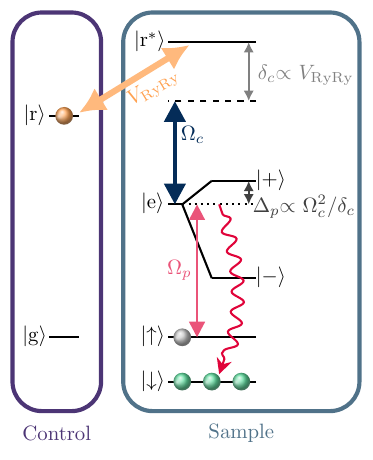}
            \label{sub_fig:scheme_on}
        \end{subfigure}
        \vspace*{-10pt}
        \caption{Level scheme in the two-level picture. We denote the atom which hosts the Rydberg excitation as control atom and refer to all other atoms as sample atoms.
        (a) Without any Rydberg excitation present, strong coupling (blue) splits the intermediate state and increases $\Dp$ of the probe field (red).
        Optical pumping to $\gaux$ (light red) is inhibited.
        (b) In presence of  $\ryd$, $\vint$ (orange) detunes $\rydaux$ and reduces the detuning $\Dp$ of the intermediate state. Optical pumping into $\gaux$ (red) is enhanced.}
        \label{fig:ati_scheme}
    \end{figure}

    Fig.\,\ref{fig:ati_scheme} exemplifies the imaging protocol on the basis of the two-level approximation. Due to the strong coupling to the Rydberg state, the intermediate state $\inter$ splits into two eigenstates $\intn$ and $\intp$, whose energy shift is given by the Rabi frequency of the coupling laser $\Oc$ and its detuning $\Dc$ \cite{PhysRev.100.703}:
    \begin{equation}
        \Delta E_{\pm}=-\frac{\hbar\Dc}{2}  \pm \frac{\hbar\sqrt{\Oc^2+\Dc^2}}{2}\,.
        \label{eq:at_full}
    \end{equation}
    In the limit of vanishing detuning, this results in the well known Autler-Townes splitting
    \begin{equation}
        \Delta E_{\pm} = \pm \frac{\hbar\Oc}{2}
        \label{eq:at_res}
    \end{equation} 
    and the atoms in the sample witness an effective detuning
    \begin{equation}
        \left|\Dp\right|=\frac{\Oc}{2}
        \label{eq:dp_res}
    \end{equation}
    on the probing transition $\g\rightarrow\inter$.
    For large $\Oc$, the initial state of the atom is preserved as scattering on the probe transition is strongly suppressed (Fig.\,\ref{sub_fig:scheme_off}).
        If, however, the control atom is in state $\ryd$, the situation is different. 
    Due to their huge polarizability, Rydberg atoms experience a strong dipole-dipole or van der Waals interaction $\vint$, which shifts the energy of $\rydaux$ \cite{PhysRevA.75.032712, Singer_2005, Comparat:10} and increases the detuning $\Dc\propto\vint$.
    For $\Dc\gg\Oc$, Eq.\,\eqref{eq:at_full} is approximated by
    \begin{equation}
        \Delta E_{\pm}=-\frac{\hbar\Dc}{2}\pm\left(\frac{\hbar\Dc}{2}+\frac{\hbar\Oc^2}{4\Dc}\right),
        \label{eq:at_offres}
    \end{equation}
    resulting in a small detuning 
    \begin{equation}
        \left|\Dp\right|=\frac{\Oc^2}{4\Dc} \ll \frac{\Oc}{2}
        \label{eq:dp_offres}
    \end{equation}
    for one of the eigenstates. Consequently, the transfer to $\gaux$ is enhanced (Fig.\,\ref{sub_fig:scheme_on}).

    While this simplified two-level consideration explains the basic principle, the multi-level structure of the atom leads to a much more involved situation as multiple couplings and competing interactions have to be considered \cite{Bender_2024}.
    Furthermore, the energy shift due to $\vint$ leading to the coupling laser detuning $\Dc$ cannot always be assumed to be independent of the Rabi coupling, as for sufficiently strong $\Oc$ both interactions compete with each other, leading to modifications of the Rydberg-Rydberg potential.
    For a more in-detail calculation bases on the simplified two-level model, we refer to Appendix\,1.

    \section{Full model}
    \label{sec:model}
    
    \subsection*{Hamiltonian and chosen basis sets}
    \label{sec:hamiltonian}
    Taking into account these aspects, we extend the discussion to the more realistic case of a multi-level system with respect to the intermediate state and the Rydberg states.
    To this end, we introduce all interactions and the appropriate basis as a starting point for the diagonalization of the system Hamiltonian.

\begin{figure}[ht]
        \captionsetup[subfigure]{oneside,margin={0.5cm,0cm}}
        \vspace*{10pt}
        \centering
        \begin{subfigure}[t]{0.15\textwidth}
            \caption{}
            \centering
            \includegraphics[width=\textwidth]{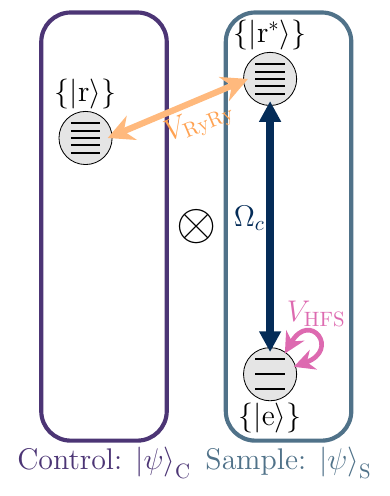}
            \label{sub_fig:multi_level_ext}
        \end{subfigure}
        \begin{subfigure}[t]{0.3\textwidth}
            \caption{}
            \centering
            \includegraphics[width=0.8\textwidth]{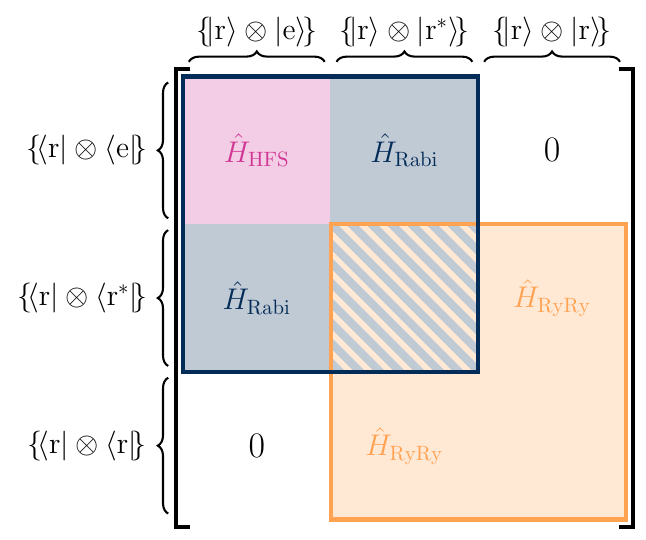}
            \label{sub_fig:ati_hamiltonian}
        \end{subfigure}
        \caption{(a) Multi-level extension of the system. 
        The formerly singular states in Fig.\,\ref{fig:ati_scheme} comprise whole manifolds. 
        $\ve$ couples the states in $\{\inter\!\}$. 
        Assuming weak probing, the ground states can be excluded. 
    (b) Structure of $\Hati$ according to Eq.\,\eqref{eq:ati_ham} with its different components acting on distinct subsets of the basis.}
        \label{fig:model}
    \end{figure}

    As we consider the probing to be weak compared to all other interactions and couplings, we don't model it but rather include the probing of the states $\g$ and $\gaux$ in a second step as a perturbation.
    A sketch of the level system is shown in Fig.\,\ref{sub_fig:multi_level_ext}.
    We include two Rydberg manifolds $\left\{\rydaux\!\right\}$ and $\left\{\ryd\!\right\}$  to realistically model the Rydberg-Rydberg interaction $\vint$.
    In addition, all states within $\left\{\inter\!\right\}$ and their internal interaction $\ve$ are included as well. 
    Finally, we neglect multi-particle effects between the atoms in the sample, which go beyond the Rydberg blockade physics.
    It is now sufficient to consider only the control atom and one additional atom, which is representative for each atom in the ensemble. We therefore restrict the Hilbert space to a two-particle product basis
    \begin{equation}
        \base=\psiq\otimes\,\psis\,,
        \label{eq:prod_basis_raw}
    \end{equation}
    where $\psiq(\psis)$ represents the state of the control (sample) atom. As the control atom is either in the Rydberg state $\ryd$ or does not play a role, the intermediate state manifold $\{\inter\!\}$ is only taken into account for the sample atom.

    With these assumptions, the Hamiltonian $\Hati$  consists of three terms, describing the involved couplings:
    
    \begin{equation}
        \Hati\left(\sep\right)=\Hryd\left(\sep\right)+\,\Hrabi+\,\Hhfs\,.
        \label{eq:ati_ham}
    \end{equation} 
    
    The Rydberg-Rydberg Hamiltonian $\Hryd\left(\sep\right)$ is based on the dipole-dipole interaction $\vint$, which couples the Rydberg states $\left\{\ryd\!\right\}$ and  $\left\{\rydaux\!\right\}$ of the control and the ensemble atom, respectively. It explicitly depends on the separation $\sep$ of the two atoms. In the following, we drop this dependence in the notation for the sake of simplicity. We, however, point out it's importance when considering the finite extent of the sample and the relative distances of the atoms.
    We set up the Hamiltonian following Ref.\,\cite{PhysRev.166.376}, including only dipolar interactions. 
    As for the dipole-dipole interaction, the fine structure basis of both atoms is a good basis, defined by quantum numbers $n, l, j$, $m_j$ and $s$:
    \begin{equation}
            \base=\psiq\otimes\,\psis=\psiqfs\otimes\,\psisfs.
            \label{eq:prod_basis_fs}
    \end{equation}
    
    In the same basis, the coupling between $\left\{\rydb\!\right\}$ and $\left\{\inter\!\right\}$ is described by the Rabi Hamiltonian $\Hrabi$, whose expectation value is given by $\Oc=\frac{1}{\hbar}\braket{\mathrm{r}^{(*)}|\,\dipoleop\cdot\hat{\mathbf{E}}\,|\mathrm{e}}$, where $\dipoleop$ is the dipole operator and $\hat{\mathbf{E}}$ the electric field of the coupling light.
    The Hamiltonian separates
    \begin{equation}
        \Hrabi=\Hrabi[, \mathrm{C}]\otimes\,\Hrabi[, \mathrm{S}]
        \label{eq:Hrabi_both}
    \end{equation}
    and if $\left\{\rydaux\!\right\}$ and $\left\{\ryd\!\right\}$ are energetically very different states, the coupling only acts on the sample atom and we can rewrite 
    \begin{equation}
        \Hrabi=\ident[\mathrm{C}]\otimes\,\Hrabi[, \mathrm{S}].
        \label{eq:Hrabi_sep}
    \end{equation}    
    
    While the hyperfine interaction for Rydberg states can be neglected \cite{PhysRevA.67.052502, PhysRevA.106.052810, B.S.:22}, it has to be included in the manifold $\left\{\inter\!\right\}$ as for lower lying states the associated energy shifts are comparable to $\Oc$. 
    The hyperfine structure Hamiltonian
    \begin{equation}
        \Hhfs=\ident[\mathrm{C}]\otimes\,\mathrm{A}_\mathrm{HFS}^{\{\inter\}}\,\mathbf{\hat{I}}\cdot\mathbf{\hat{J}}
        \label{eq:Hhfs_sep}
    \end{equation} 
    again only acts on the sample atom $\psis$.
    To this end, the nuclear spin $I$ and its projection $m_I$ of the sample atom need to be included, extending Eq.\,\eqref{eq:prod_basis_fs} to
    \begin{equation}
        \base=\psiqfs\otimes\,\psisfss.
        \label{eq:prod_basis_fss}
    \end{equation} 
    In this basis, the Hamiltonian $\Hati(R)$ takes the form presented in Fig.\,\ref{sub_fig:ati_hamiltonian} and is diagonalized to obtain the level structure that can be addressed by the probing laser to perform optical pumping.

    \subsection*{Probing}
    \label{sec:probing}

    As a second step, we now consider the scattering of the (weak) probing laser on the level system resulting from the above diagonalization.
    The primary quantity of interest in this process for our proposal is the sample transfer probability $\transfer$. It is defined as the probability of transferring an atom of the sample from $\g$ to $\gaux$ and will in the following be calculated.

    To simplify the calculation, we first transform the eigenstates of $\Hati$ to the hyperfine structure as we naturally describe the ground states in the basis $\psig=\ket{n, j, l, F, m_F}$.
    The manifolds $\left\{\ryd\!\right\}$ and $\left\{\rydaux\!\right\}$ do not need to be transformed.

    Since we are only interested in the ground state population of the sample atoms, irrespective of the control atom, we need to trace out the latter state.
    To this end, we set up the density matrix \mbox{$\dmatrix[\mathrm{CS}]^{\left(\varphi\right)}=\egbase[\varphi]\egbasebra[\varphi]$} for each eigenstate $\egbase[\varphi]$ resulting from the diagonalization and take the partial trace over the control atom:
    \begin{equation}
        \dmatrix[\mathrm{S}]^{\left(\varphi\right)}=\mathrm{tr}_\mathrm{C}\left(\dmatrix[\mathrm{CS}]^{\left(\varphi\right)}\right).
        \label{eq:part_trace}
    \end{equation}
    Diagonalizing the density matrix provides a basis set $\left\{\ket{\Theta_\vartheta}\!\right\}$ with eigenvalues $\left\{\theta_\vartheta\right\}$, in which the coupling of the probe field can be treated conveniently.
  
    For each ground state $g \equiv \psig[g]$ and basis state $\vartheta \equiv \dmes[\vartheta]$, we determine the scattering rate according to
    \begin{equation}
        \scattot^{\left(g,\vartheta\right)}\,=\,\frac{\Gamma}{2}\,\frac{2\left( \Omega^{\left(g,\vartheta\right)}\!/\Gamma\right)^2}{1+2\left( \Omega^{\left(g,\vartheta\right)}\!/\Gamma\right)^2+\left(2\Dp/\Gamma\right)^2}\,.
        \label{eq:ev_scat_rate}
    \end{equation}
    Here, $\Omega^{\left(g,\vartheta\right)}\,=\,{}_\mathrm{G}\!\braket{\psi_{g}|\,\dipoleop\cdot\hat{\mathbf{E}}\,|\Theta_\vartheta} \label{eq:ev_rabi_coupling} $ is the Rabi coupling between the states $g$ and $\vartheta$, $\Gamma$ is the natural linewidth of the $\g\rightarrow\inter$ transition and the effective probe laser detuning $\Dp$ is given by the difference in the photon energy and the energy of the eigenstate $\egbase[\varphi]$.
    Summing up the contributions of the different $\dmes[\vartheta]$ results in
    \begin{equation}
        \scattot^{\left(g, \varphi\right)}\,=\,\sum_{\vartheta}\left|\theta_\vartheta\right|^2 \scattot^{\left(g,\vartheta\right)}.
        \label{eq:esgs_scat_rate}
    \end{equation}
    In the limit of weak probing, the scattering rates of the individual eigenstates $\egbase[\varphi]$ for a given ground state $\psig[g]$ can be summed up in a single effective scattering rate
    \begin{equation}
        \scattot^{\left(g\right)}\,=\,\sum_{\varphi}\scattot^{\left(g, \varphi\right)}\,.
        \label{eq:gs_scat_rate}
    \end{equation}
    The above expression is the basis for the calculation of the transfer probability. Note again, that it implicitly depends on $\sep$.   
   
    The number of photons scattered by an atom is then given by 
    \begin{equation}
        \nscat\,=\,\scattot^{\left(g\right)}\cdot\tp\,,
        \label{eq:n_scat_g}
    \end{equation}
    
    where $\tp$ is the duration of the probe pulse. To access the transfer probability we define the branching ratio $\branch$, which describes the probability to end up in $\gaux$ after one scattering event.
    After the probing time $\tp$, we find
    
    \begin{equation}
        \transferr^{\left(g\right)}\,=\,1-\left(1-\branch[]\right)^{\nscat}\,,
        \label{eq:state_transfer_rg}
    \end{equation}
   for a given initial state $\g$, with probability $p^{\left(g\right)}$ and \mbox{$\sum_{g}p^{\left(g\right)}=1$}. The weighted transfer probability is then given by
    \begin{equation}
        \transferr\,=\,\sum_{g}p^{\left(g\right)}\transferr^{\left(g\right)}\,.
        \label{eq:state_transfer_r}
    \end{equation}
    Note that due to the $\sep$-dependence of the eigenstates, this probability is still dependent on the internuclear separation of control and sample atom.
    Therefore, the transfer probability of the whole sample
    \begin{equation}
        \transfer\,=\,\int_{0}^{\infty}\dens\left(\sep\right)\transferr\left(\sep\right)\,\mathrm{d}\sep.
        \label{eq:state_transfer}
    \end{equation}
    takes this fact into account by introducing the internuclear distance probability density $\dens$.
    It depends on the trapping potential and the temperature of the sample. In the following, we assume a harmonic trapping potential and temperature of $50\,\mathrm{\mu K}$ which is typical after a dark MOT loading into the tweezer potential. Further details on the sample properties can be found in the Appendix\,3.

    \section{Numerical simulations}
    \label{sec:numerics}
    
    The fidelity of the imaging protocol is determined by the difference between the transfer probability Eq.\,\eqref{eq:state_transfer} with and without Rydberg excitation present in the ensemble.   
    In the following we will assess this Rydberg imaging fidelity in a realistic scenario and investigate its dependence on the systems parameters. 
 
    \subsection*{Choice of involved atomic states}
    \label{sec:states}
    
    The presented imaging scheme is very versatile and can be tailored to different needs.
    Finding optimal parameters, on the other hand, is specific to the concrete application and depends on many aspects like the atomic species, the experimental implementation, the available laser systems, the number of atoms, the size of the sample, and the purpose of the experiment. 
 
    Here, we focus on mesoscopic ensembles of rubidium atoms in optical tweezers. 
    We consider the two hyperfine ground state manifolds of rubidium $\left\{\g\!\right\}=\gsnum[,\,m_F]$ and $\left\{\gaux\!\right\}=\gsnumaux[,\,m_F]$.
    As intermediate state, we chose the 6P state, $\left\{\inter\right\}=\intnum[;\,F,\,m_f]$ state. Within this manifold, we near resonantly address the transition $\gsnum[]\rightarrow\intnum[; F=2]$. 
    For a given Rydberg state $\ryd$, which is to be detected, the auxiliary state $\rydaux$ needs to be a different Rydberg state in order to prevent a fast map down and loss of the excitation due to the light coupling.
    Furthermore, in order to ensure strong interactions across the whole sample, we make use of Förster resonances \cite{https://doi.org/10.1002/andp.19484370105, PhysRevA.77.032723, Wu:23}, substantially increasing $\vint$ while maintaining large Rabi couplings.
    Specifically, we choose the
    \begin{equation*}
        \rydnum\otimes\rydnumaux[]\leftrightarrow\rydfo[]\otimes\rydfo[]
    \end{equation*}
    resonance with a Förster defect of only $\Delta_\mathrm{F}=2\pi\times 4.3\,\mathrm{MHz}$.
    While the proposed imaging scheme generally works with any pair of Rydberg states, it obviously benefits from the strong enhancement of the interaction at the Förster resonance.
    For a detailed discussion of the state and parameter choice see Appendix\,4.
    
    To sufficiently describe the Rydberg interaction, several other close by Rydberg states need to be includes, as due to their energetic proximity, they still have a non vanishing influence on the shifts of the Rydberg states of interest (see Appendix\,2).

\begin{figure}[h]
        \centering
        \includegraphics[width=0.33\textwidth]{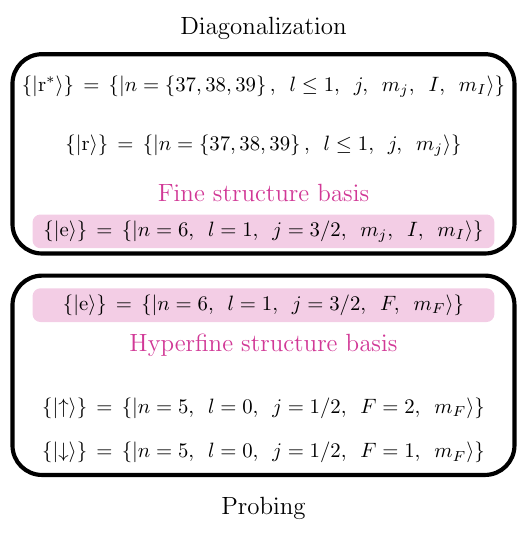}
        \caption{Basis sets to evaluate $\Hati$ (top) in fine structure basis and the probing (bottom) in hyperfine structure basis.
        }
        \label{fig:num_states}
    \end{figure}

    An overview of the basis states for the following numerical simulation is given in Fig.\,\ref{fig:num_states}.

    \subsection*{Experimental parameters}
    \label{sec:parameters}

    Besides the number of included basis states, there are many more parameters to be chosen. 
    
    While not being exhaustive by any means, we chose in the following experimental settings which are frequently encountered in experiments with optical tweezers and allow us at the same time to illustrate the basic dependencies of the imaging protocol.
    Regarding the light field, the polarization $q_c$ and detuning $\Dc$ with respect to the $\intnum[; F=2]\rightarrow\rydnumaux[]$ transition are kept fixed throughout the simulations.
    Since the coupling strength $\Oc$ varies for different $\intnum[;\,F=2,\,m_F]$ states, we use the smallest of those couplings as a reference to denote the coupling strength in the different analyzed situations.
    
    We consider only $q_p=+1$ polarization in the probing transition and fix the probing time to $t_p=15\,\si{\micro\second}$ which corresponds to half the lifetime of the $\rydnum[]$ state.

    To optimize the imaging scheme, our primary observable of interest is the relative transfer probability, i.e. the transfer ratio $\ratio=\transfer/\transfer_\mathrm{no Ryd}$, where $\transfer_\mathrm{no Ryd}$ is the sample transfer probability when the control atom is not in the Rydberg state.
    However, since we are dealing with relatively small particle numbers, we also need to make sure that the absolute number of transferred atoms Eq.\,\eqref{eq:state_transfer} remains detectable.
    From these quantities, we can determine the detection fidelity.
    The chosen parameters are listed in Tab.\,\ref{tab:params}.
    For the initial population probabilities $p_\mathrm{G}$ in Eq.\,\eqref{eq:state_transfer_r} we assume a uniform distribution between all Zeeman sublevels.
    To correctly account for the competition between the Rydberg-Rydberg interaction and the strong Rabi coupling, we set up and diagonalize the Hamiltonian $\Hati$ for varying internuclear separations $\sep$ and Rabi couplings $\Oc$.

    \begin{table}
        \caption{
            Parameters for the diagonalization (top left), probing (top right) and sample (bottom).
            Fixed parameters are given with respective value.
            }
        \begin{center}
            \begin{tabular}{| l | c | c | l | c | c |}
                \hline
                \multicolumn{3}{| c |}{\centering\textbf{Diagonalization}} & \multicolumn{3}{| c |}{\centering\textbf{Probing}} \\
                \hline\hline
                coupling strength & $\Oc$ & varied & probing strength & $\op$ & $2\pi\times 1\si{\mega\hertz}$ \\
                polarization & $q_c$ & $-1$ & polarization & $q_p$ & $+1$ \\
                detuning & $\Dc$ & $0\,\si{\mega\hertz}$ & detuning & $\Dp$ & varied \\
                \hline\hline
                \multicolumn{6}{| c |}{\centering\textbf{Additional}}\\
                \hline\hline
                tweezer waist  & $w_0$ & $2\,\si{\micro\meter}$ & temperature & $T$ & $50\,\si{\micro\kelvin}$ \\
                trap depth & $U_0$ & $1\,\si{\milli\kelvin}$ & pulse duration & $t_p$ & $15\,\si{\micro\second}$ \\
                \hline
            \end{tabular}
        \end{center}
        \label{tab:params}
    \end{table}

    \section{Results}
    \label{sec:results}

    We now present the results of the numerical simulations for the diagonalization, the transfer probability and the transfer ratio. 

    \subsection*{Diagonalization}
    \label{sec:diagonalization_results}

    \begin{figure}
        \centering
        \includegraphics[width=0.45\textwidth]{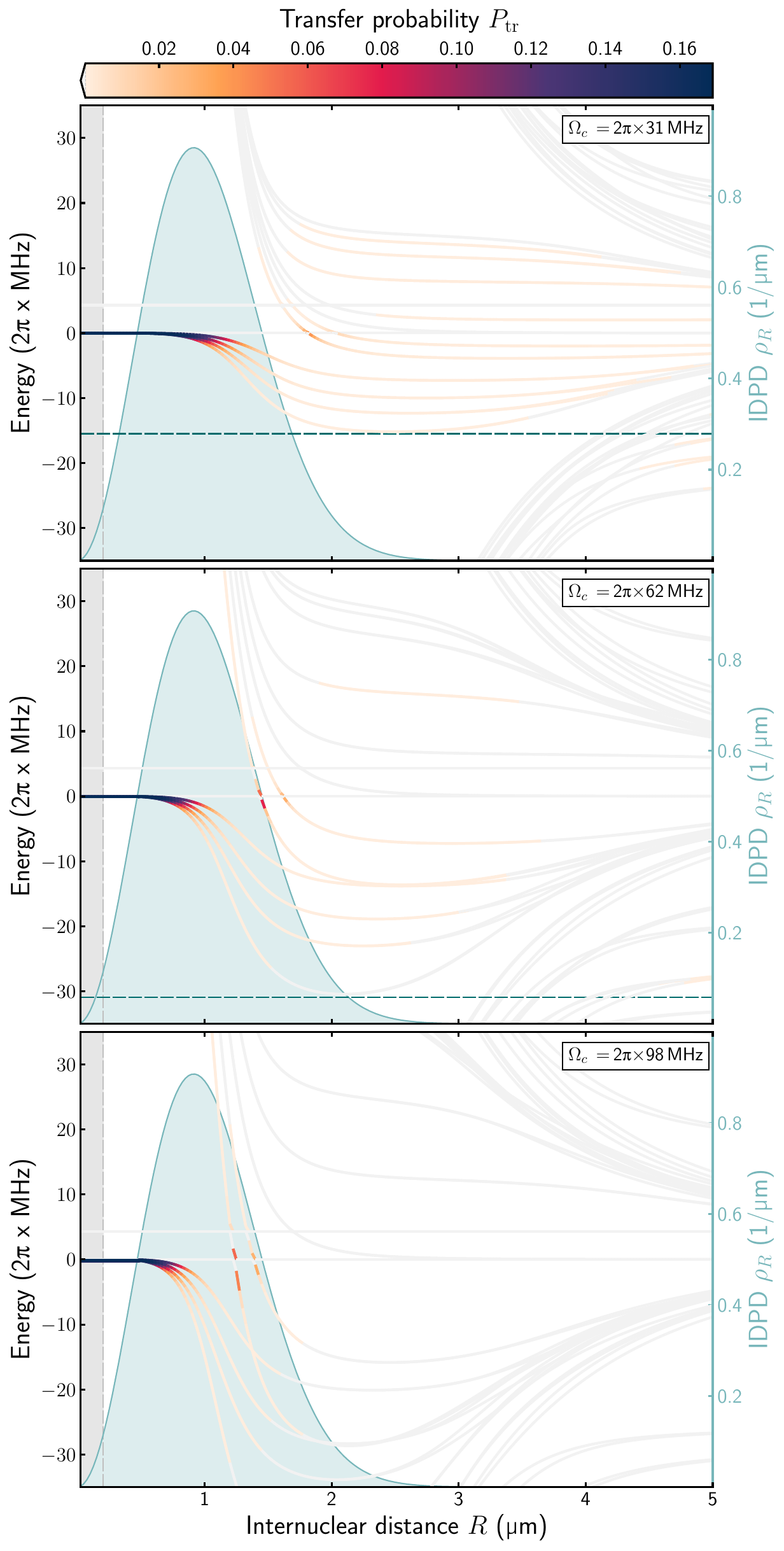}
        \caption{Eigenenergies of the ATI system for increasing $\Oc$ (top to bottom).
        The on-resonance transfer probability $\transferr$ is given as a color code, values $\transferr\leq 0.01\transferr^\mathrm{max}$ are depicted in light gray.
        In the gray shaded area, the eigenenergies and state compositions are assumed to be constant, the petrol curve is the internuclear distance probability.
        The transfer probability is maximized for small internuclear distances where $\vint$ dominates the energy scales. 
        In the intermediate regime the coupling starts to split the formerly nearly degenerate eigenstates.
        For large $\sep$, the Rabi coupling is the main contributor and transfer is strongly suppressed by the Autler-Townes splitting.
        The reference energy is set to the $\ket{39S_{1/2};\,6P_{3/2},\,F=2}$ state.}
        \label{fig:spectra}
    \end{figure}

    In Fig.\,\ref{fig:spectra} we show the distance dependent eigenenergies of the system.
    A multitude of lines at different energies is present in the spectrum.
    To concentrate on the later on important ones, we already color-coded the transfer probabilities $\transferr$ (see Section\,\ref{sec:probing_results}) where we can identify three different regimes.
    We find the maximal $\transferr$ in the regime of small distances $\sep < 1\si{\micro\meter}$ for all coupling strengths $\Oc$. 
    This is to be expected from the simple few-level model as in this range, the Rydberg-Rydberg interaction is dominant.
    The $\rydnumaux[]$ state gets shifted out of resonance for the coupling field, leading to a breakdown of the Autler-Townes condition.
    In the intermediate distance regime ($1\si{\micro\meter}\lesssim\sep\lesssim 4\si{\micro\meter}$), $\Oc$ becomes comparable to $\vint$, resulting in a splitting of the formerly nearly degenerate eigenstates caused by the differing dipole matrix elements of the dominantly admixed $\intnum[;\,F,\,m_F]$ states according to their Clebsch-Gordan decomposition.
    States with larger $m_F$ experience larger shifts, as the polarization $q_c=-1$ favors the $\intnum[;\,m_j=+\sfrac{3}{2}]$ component.
    The relative splitting increases with $\Oc$, as well as the eigenenergy variation in a given distance interval.
    Additionally, in this regime, eigenstates approach the reference energy from above and show a largely increased transfer probability in the vicinity of zero energy. 
    Those states carry a large admixture of $\intnum[;\,F=3,\,m_F=3]$ character and are shifted into resonance with the probing laser by the coupling light field when the Rydberg-interaction induced shift decreases with increasing $\sep$.
    With increasing coupling strength, this isolated increase in transfer probability moves towards smaller $\sep$, as the splitting relative to the Rydberg-shifted $F$ state by the Rabi coupling increases as well.

    In the regime of large $\sep$, the transfer probability significantly decreases and the states with a residual $\transferr$ can be classified into two types.
    The ones which form the continuation of the previously discussed states approaching $E=0\si{\mega\hertz}$, for whom $\transferr$ decreases with increasing $\Oc$. 
    Those states contain an ever decreasing admixture of $\left\{\inter\!\right\}$ but are closer to resonance than the second type which is only visible for large $\sep$.
    These states are situated at roughly $-\frac{\hbar\Oc}{2}$, the bare Autler-Townes energy shift according to Eq.\,\eqref{eq:dp_res}.
    In this distance regime, the Rabi coupling is dominant, the multiple lines visible corresponding to different coupling strengths of the $\intnum[; F=2]\rightarrow\rydnumaux[]$ transitions.
    The states possess large $\left\{\inter\!\right\}$ admixtures but are far detuned, thereby decreasing $\transferr$.
    With $\Oc$, the Autler-Townes splitting increases further suppressing the transfer.

    Looking at the internuclear distance probability density (IDPD), we find sufficient transfer in the regions where most of the atoms are situated.
    The regime in the absence of the control excitation becomes visible at large $\sep$, where the Rydberg-Rydberg interaction becomes negligible.

    \subsection*{Probing}
    \label{sec:probing_results}
    As outline above, the sample transfer probability $\transfer$ and the transfer ratio $\ratio$ are the main quantities to assess the fidelity of the imaging protocol.
    To this end, we focus on the probe laser detuning and the coupling laser strength in order to optimize the imaging parameters.

    \begin{figure}[ht]
        \centering
        \includegraphics[width=0.45\textwidth]{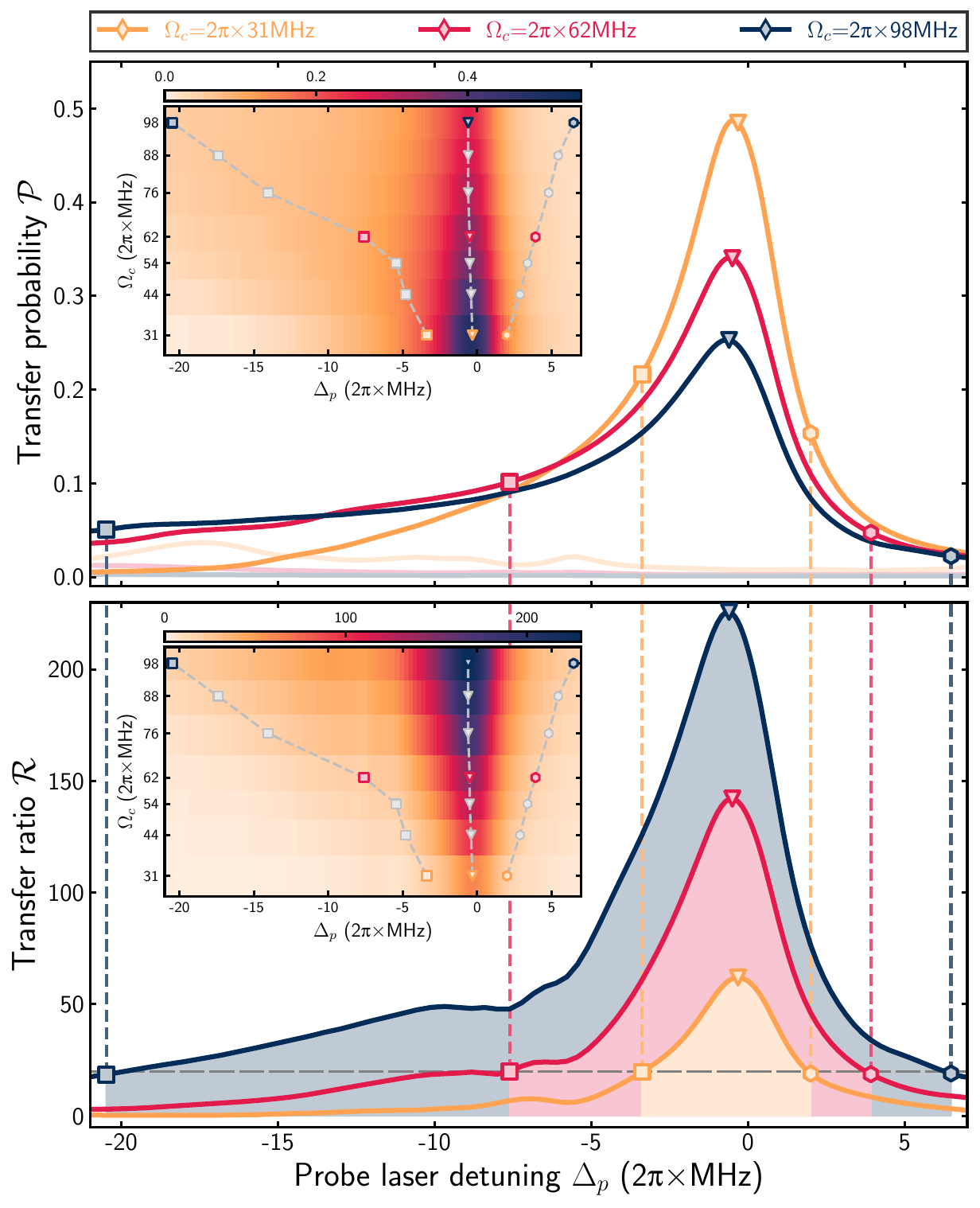}
        \caption{Transfer probability and transfer ratio for varying $\Dp$ and $\Oc$, insets showing the associated quantity for a wider range $\Oc$.
        (Top) Up to $50\,\si{\percent}$ of atoms are transferred in the given configuration with amplitude decreasing for increasing $\Oc$. 
        For negative probe detunings, the line is significantly broadened, making for a more robust setup of the scheme.
        Semi opaque lines give the probability for control atom $\gs$.
        (Bottom) Conditions in terms of $\Oc$ invert for $\ratio$ with better contrasts achieved for larger couplings.
        Exemplarily we indicate $\ratio=20$ (dashed gray line) with detuning(s) of the intersection(s) marked in both plots.
        Color of inset markers and cuts in the main plot coincide, gray line in inset giving the mentioned intersection.
        }
        \label{fig:probing}
    \end{figure}

    Fig.\,\ref{fig:probing} shows the transfer probabilities and transfer ratios for different probe laser detuning $\Dp$ and a selection of different coupling strengths.
    For all $\Oc$ we find a sufficient transfer probability, reaching up to $50\,\si{\percent}$ for the smallest coupling strength, with maximal values at small negative detunings.
    While decreasing on the scale of the natural linewidth $\Gamma$ for positive detunings, the spectrum is significantly broadened for negative $\Dp$, with the broadening becoming more pronounced for increasing $\Oc$.
    This behavior can be understood from the eigenenergies in Fig.\,\ref{fig:spectra} where,
    for positive detunings, no eigenstates are in the energetic vicinity and the probe laser off-resonantly probes the states at $E=0\,\si{\mega\hertz}$, leading to a decrease of the scattering rate on the scale of $\Gamma$.
    For negative detunings, however, the eigenstates can still be resonantly probed at larger $\sep$. 
    Since at this point, the coupling leads to a splitting of the curves, not all these eigenenergy curves can be addressed simultaneously, hence $\transfer$ decreases.
    The states within the fan with larger energy shift can on the other hand efficiently be probed for larger absolute $\Dp$, causing the broadening of the spectral line for negative detunings.
    This circumstance makes the system robust against fluctuations in the laser frequency and residual light shifts. This is in contrast to EIT imaging schemes, where the spectral narrowness of the dark state requires a hight control of the involved line shifts and frequency drifts \cite{PhysRevLett.108.013002}.
    
    As shown in the lower part of Fig.\,\ref{fig:probing}, the transfer ratio increases with the coupling strength and thus shows the inverse behavior compared to the transfer probability just discussed.
    This can be understood in the simple two-level picture where the Autler-Townes splitting increases linearly with $\Oc$ and thus the scattering rate decreases quadratically according to Eq.\,\eqref{eq:ev_scat_rate}.
    It is especially in the intermediate distance regime, where the effect of the coupling increases the effective detuning and thus changes the scattering rate.
    Since a significant fraction of particles have distances in this regime, this leads to a less pronounced suppression of the scattering rate, increasing $\ratio$ for large coupling strengths.

    The opposite scaling of the performance indicators $\transfer$ and $\ratio$ requires to find an optimal trade-off that depends on the application and the experimental setup at hand.
    For larger atom numbers $N$ for example, a smaller amount of transferred particles can still be sufficient to detect the excitation if few or single atom detection sensitivity is achieved.
    In this case, increasing $\Oc$ allows for a higher contrast detection for a wider range of detunings.

    \subsection*{Detection fidelities}
    \label{sec:fidelity}

    \begin{figure}[ht]
        \centering
        \includegraphics[width=0.45\textwidth]{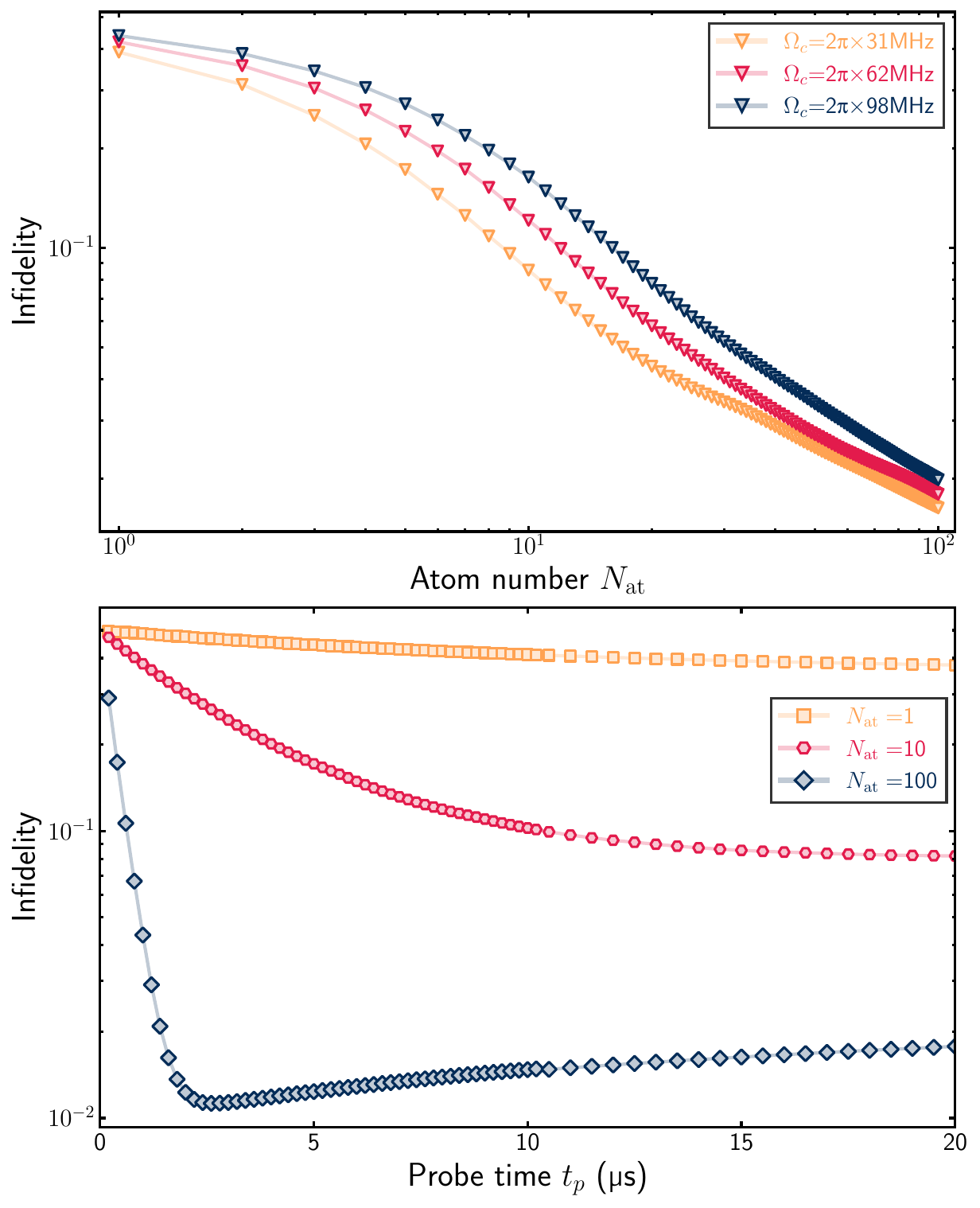}
        \caption{(Top) Detection infidelity for different $\Oc$ and $\Nato$ for a probe time $t_p=15\,\si{\micro\second}$.
        For increasing atom number, the infidelity decreases.
        As $\transfer$ is better for lower coupling with $\ratio$ still being large, the infidelity decreases with $\Oc$ at the optimal point (see Fig.\,\ref{fig:probing}).
        (Bottom) Infidelity for different $t_p$ at $\Oc=2\pi\times 31\,\si{\mega\hertz}$ .
        Due to the decay of the control excitation, there is an optimal probe time depending on $\Nato$, shifting towards shorter times for increasing atom number.
        }
        \label{fig:infidelity}
    \end{figure}

    To benchmark the ATI protocol, we calculate the detection infidelities at the parameters marked by triangles in Fig.\,\ref{fig:probing} and vary the number of sample atoms $\Nato$ and the duration of the probe pulse $t_p$.
    To this end, we include the Poissonian distribution of both, the number of transferred atoms and the number of photons scattered by the probing light.
    The Poisson distribution for the number of transferred atoms is characterized by a mean value depending on $\transfer$ and $\Nato$.
    For the subsequent detection, we assume a fluorescence imaging protocol with $\bar{n}_\mathrm{ph}=10$ detected photons per atom.
    This relatively small number of detected photons accounts for the loss of atoms in a dense sample due to light-assisted collisions \cite{PhysRevA.85.062708} during the fluorescence image and can be considered a conservative assumption to find a lower bound for the fidelities. 
    It is furthermore crucial to take into account the decay of the control excitation that can happen at random times within the duration of the probe pulse.
    A detailed discussion of the modelling of the fidelity is found in Appendix\,5.
    
    The obtained scaling of the infidelities are shown in Fig.\,\ref{fig:infidelity} for varied coupling strengths and atom numbers.
    It reveals that the infidelity decreases for larger samples.
    This is expected since the probing is a single particle effect and thus the absolute number of transferred atoms scales with the number of atoms in the ensemble, irrespective of the control atom being in the Rydberg state or not.
    Analyzing the scaling with the coupling strength, we find that moderate couplings (which also maximize $\transfer$) are favorable as they show generally smaller infidelities,
    The coupling must, however, be chosen large enough that the transfer ratio $\ratio$ remains  sufficiently large.
    For an easily discriminable transfer ratio $\ratio > 10$ this is given for all parameters analyzed in Fig.\,\ref{fig:infidelity}.
    
    The fidelities can even be improved by using adequately chosen probe pulse durations as depicted in the lower panel of Fig.\,\ref{fig:infidelity}.
    In general longer probe times allow for more scattering events and thus more atoms to be transferred.
    On the other hand longer probe times also make a decay of the Rydberg atom during the probe pulse more likely.
    These contrary effects lead to the existence of an optimal probing time that depends on the coupling strength and the number of atoms in the ensemble.
    In accordance with the previous results, the optimal probe time gets reduced for increasing $\Nato$, to the point where pulses of only a few $\si{\micro\second}$ are ideally suited to map the state of the control atom to the surrounding sample, underlining the potential of the protocol to very rapidly measure and store the state of a Rydberg atom.

    In comparison to Rydberg imaging techniques based on the breakdown of the EIT condition \cite{PhysRevLett.127.050501}, the presented ATI scheme acts on similar timescales but surpasses the achievable fidelities already at relatively small sample sizes of only $\Nato >10$.
    To obtain the same imaging fidelity the EIT-based protocol requires much larger atom numbers $\Nato\approx400$ and suffers from spectroscopically narrow EIT feature.

    \section{Discussion}
    \label{sec:discussion}
    Exploiting the competition of Rydberg-Rydberg interaction and external Rabi coupling, our novell detection scheme for individual Rydberg excitations encodes the state of the control atom in the surrounding sample within only a few $\si{\micro\second}$.
    At this time the information is stored in a stable state and preserved for later readout.
    In addition to the swiftness of the protocol, atoms only need to scatter a few to no photons, preventing losses af sample atoms due to radiative heating or molecule formation.
    For the parameters discussed within this proposal, we show robustness against external fluctuations offsetting the probe laser detuning such as residual light shifts.   
    The detection fidelities are promising, well surpassing state of the art EIT methods at much smaller atom numbers, making it ideally suited for small mesoscopic ensembles.
    Furthermore, no specialized detection equipment such as APDs with single photon sensitivity is required but the concept is readily implemented in systems with fluorescence imaging setups.
    Our scheme is, however, not limited to fluorescence imaging in tweezer setups but can be combined with every state-resolved imaging of high density samples with a resolution on the length scale of $\qty{1}{}-\qty{2}{\micro\meter}$.
    This enables its application for example in quantum gas microscopes \cite{Gross2021} where in an optical lattice with $\lambda_\mathrm{lattice} = \qty{1064}{\nm}$ 25 atoms are located in a $\qty{2.5}{\micro\meter} \times \qty{2.5}{\micro\meter}$ region and can be image state selectively.

    The method can even be extended to multi-species systems as the Rydberg-Rydberg interaction is not limited to within a single species.
    Infidelities can be further decreased by optimizing experimentally easy accessible parameters such as the coupling strength $\Oc$.
    Increasing the principle quantum number of the involved Rydberg states, we predict an increased interaction radius due to the increasing strength of the Rydberg-Rydberg interaction.
    At this point, the control atom and sample could even be spatially separated to reduce their mutual influence when the coupling is not applied.
    All aspects show the potential of the ATI scheme for state detection in modern day tweezer platforms for quantum simulation with small mesoscopic ensembles.
    
    \FloatBarrier

        \renewcommand{\bibsection}{\section*{Supplementary References}}

        \title{Supplementary Information for\\ Fast and robust detection of single Rydberg excitations in mesoscopic ensembles}
        \makeatother

        \newcommand{\beginsupplement}{
            \setcounter{table}{0}
            \renewcommand{\tablename}{Supplementary Table}
            \setcounter{figure}{0}
            \renewcommand{\figurename}{Supplementary Figure}
        }
        \clearpage
        
        \beginsupplement
        \setcounter{section}{0}
        \section*{Supplementary Information}

        \section{Two-level approximation}
        \label{appsec:two_level_approx}
        The general principle underlying the ATI scheme can be understood from a simple two-level system with two particles(see Section\,2).
        To understand the restrictions of this simplified model and to underline the quantitative requirement of the full calculation, the transfer probabilities within this restricted approximation are evaluated in the following.

        \subsection{Rydberg energy shift}
        \label{appsec:ryd_energy_shift}
\begin{figure}[ht]
            \centering
            \includegraphics[width=0.45\textwidth]{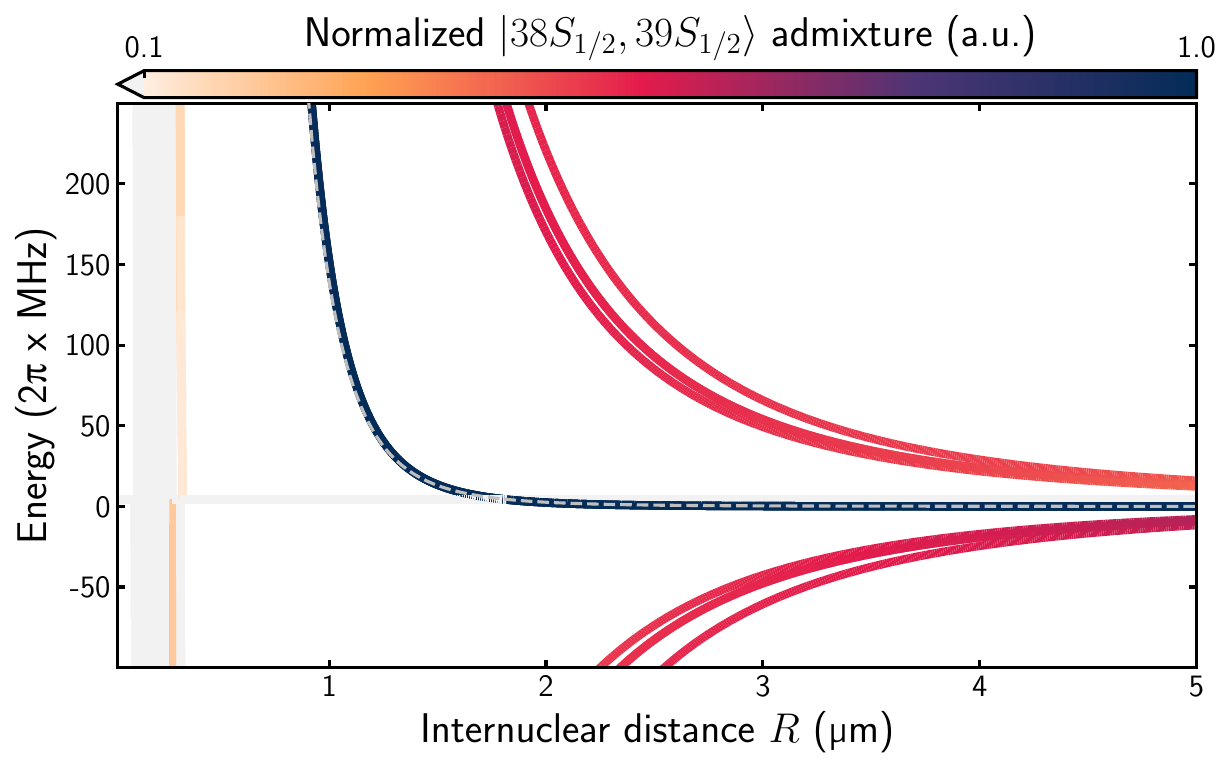}
            \caption{Eigenspectrum of the Rydberg-Rydberg interaction Hamiltonian. 
            The color code shows the normalized admixture of all $\ket{38S_{1/2}, 39S_{1/2}}$ states.
            Fitting the eigenstate with the largest admixture gives the interaction coefficients $C_3=3.8\,\si{\mega\hertz\micro\meter^3}$ and $C_6=135.2\,\si{\mega\hertz\micro\meter^6}$ (gray dashed line).
            }
            \label{fig:ryry_interaction}
        \end{figure}

        In the considered system, the eigenenergies are governed by the Rabi frequency of the coupling field $\Oc$ as well as the energy shift of the auxiliary Rydberg state $\rydaux$ which is induced by the Rydberg-Rydberg interaction and which results in an effective detuning $\Dc\left(\sep\right)$ for the coupling laser.
        We thus calculate the energy shift by diagonalizing the interaction Hamiltonian $\Hryd$ up to dipolar order \cite{PhysRev.166.376} in the pair state basis $\ket{\mathrm{r}, \mathrm{r}^*}=\ryd\otimes\rydaux$ and include all fine structure sub states of the $\ket{n=\left\{37, 38, 39\right\}, l\leq 1}$ manifolds.
        The results of the diagonalization as a function of the internuclear distance are depicted in Fig.\,\ref{fig:ryry_interaction}.
        The eigenvectors obtained from the diagonalization are used to calculate the admixture of all $\ket{38S_{1/2}, 39S_{1/2}}$ states in the new eigenstates.

        For the eigenstate with the largest admixture, the interaction coefficients $C_3$ and $C_6$ are extracted by fitting the relation
        \begin{equation}
            E\left(R\right)=\frac{C_3}{R^3}+\frac{C_6}{R^6}\,,
            \label{eq:ryry_interaction}
        \end{equation}
        to the diagonalization result.
        Exploiting the fact that the two terms in Eq.\,\eqref{eq:ryry_interaction} are dominant in different distance regimes, we make the fit more reliable by fitting the two terms individually on the respective interval.
        We use the intervals  $R_3=\left[3\si{\micro\meter}, 5\si{\micro\meter}\right]$ and $R_6=\left[0.6\si{\micro\meter}, 3\si{\micro\meter}\right)$ for the according coefficient.
        With this method we get $C_3=3.8\,\si{\mega\hertz\micro\meter^3}$ and $C_6=135.2\,\si{\mega\hertz\micro\meter^6}$ with the energy shift according to Eq.\,\eqref{eq:ryry_interaction} shown as a gray dashed line in Fig.\,\ref{fig:ryry_interaction}.

        \subsection{Transfer ratio}
\begin{figure}[ht]
            \centering
            \includegraphics[width=0.45\textwidth]{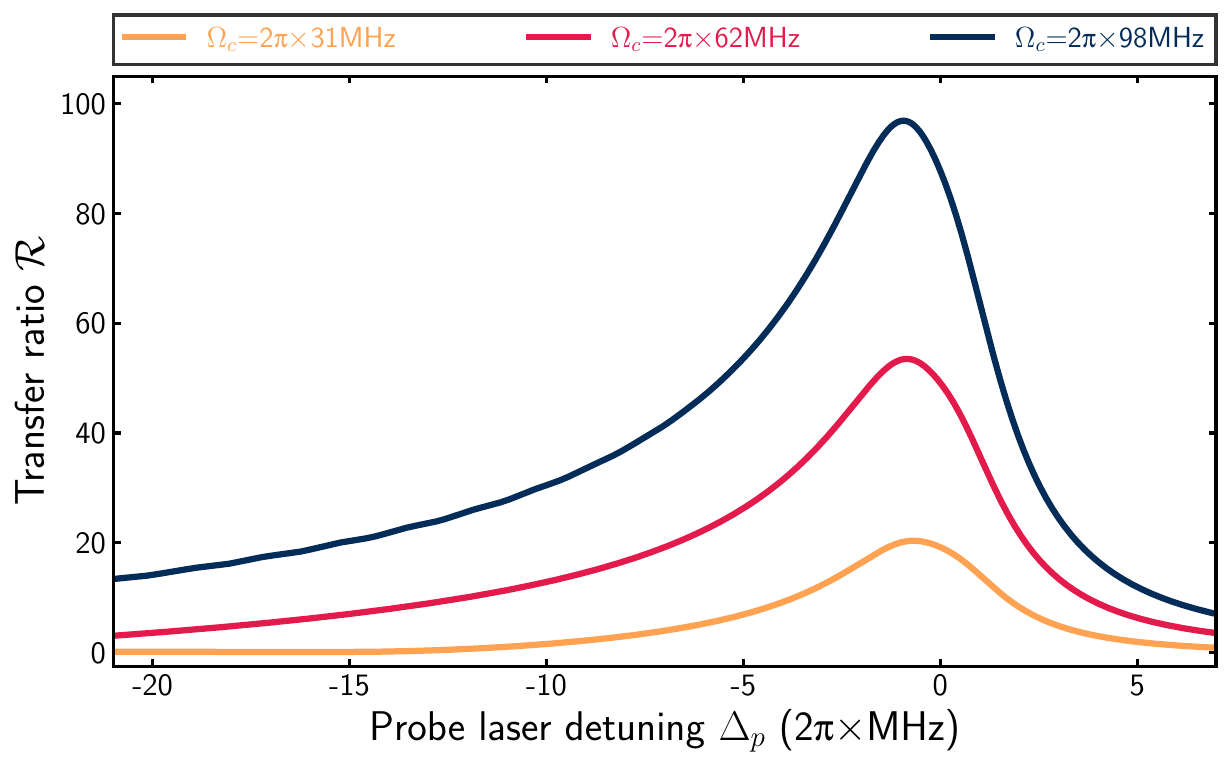}
            \caption{Transfer ratio using the simple two-level approximation for varying Rabi coupling $\Oc$.
            While the overall shape resembles the results of the full model, we see a quantitative mismatch by approximately a factor of $2$.
            }
            \label{fig:two_level_ratio}
        \end{figure}

        Taking this result, we have access to the detuning of the coupling laser $\Dc\left(\sep\right)$ and calculate the energy shift of $\inter$ according to
        \begin{equation}
            \Delta E_{\pm}\left(\sep\right)=-\frac{\hbar\Dc}{2}\pm\frac{\hbar\sqrt{\Oc^2+\Dc\left(\sep\right)^2}}{2}\,
            \label{eq:at_full_app}
        \end{equation}
        for different coupling strengths $\Oc$ in the cases with and without excitation of the control atom.
        Inserting this energy shift as an additional detuning for the probe laser
        \begin{equation}
            \Delta_{p}^{\mathrm{eff}}\left(\sep\right)=\Delta_p+\Delta E_{\pm}\left(\sep\right)
            \label{eq:effective_detuning}
        \end{equation}
        in the scattering rate solution of the two-level optical Bloch equation
        \begin{equation}
            R_\mathrm{sc}\left(\sep\right)=\frac{\Gamma}{2}\frac{2\left(\Omega_p/\Gamma\right)^2}{1+2\left(\Omega_p/\Gamma\right)^2+\left(2\Delta_{p}^{\mathrm{eff}}\left(\sep\right)/\Gamma\right)^2}
            \label{eq:scattering_rate}
        \end{equation}
        for a given probing strength $\Omega_p$, the transfer probability is determined by summing up the rates of both solutions in Eq.\,\eqref{eq:at_full_app}
        \begin{equation}
            \transferr\left(\sep\right)=1-\left(1-\branch[]\right)^{R_\mathrm{sc}\left(\sep\right)\cdot t_p}\,,
        \end{equation}
        where $t_p$ is the duration of the probe pulse and $\branch[]$ the probability to end up in the $\gaux$ state after spontaneous emission from the $\inter$ state.
        Weighting the transfer probabilities with the pair distance probability density (see Appendix\,\ref{appsec:idpd}), the sample transfer probabilities $\transfer$ and $\transfer_\mathrm{no Ryd}$ and thereby the transfer ratio $\mathcal{R}$ are finally calculated.
        For the parameters discussed in the main part, the results are shown in Fig.\,\ref{fig:two_level_ratio}.

        The broadening towards negative detunings observed in the full model (see Fig.\,6) is also visible for the simplified model, but the maximal transfer amplitude is only roughly half as high.
        Multiple effects contribute to this reduction.
        First of all, only the smallest of the Rabi couplings $\Oc$ from the full model is taken as the Rabi frequency in the simple model, effectively lowering the Autler-Townes splitting in the case without control excitation and therefore increasing the transfer in that regime.
        Secondly, we ignore all but one eigenstate in the eigenspectrum of Fig.\,\ref{fig:ryry_interaction}.
        Thereby, we ignore states states which still posses a significant admixture of the base state and where the system can scatter from.
        The additional additional scattering channels increase the transfer probability in the case of an excited control atom.
        Thirdly and most importantly, in the two-level simplified model, the Rabi coupling is always seen as a perturbation to the Rydberg-Rydberg interaction, as we first calculate the coupling laser detuning and insert it in Eq.\,\eqref{eq:at_full_app} in a second step. 
        Looking at Fig.\,\ref{fig:ryry_interaction} and the coupling strengths in Fig.\,\ref{fig:two_level_ratio}, this is obviously not reasonable, as for most $\sep$, the Rabi coupling is in the order or even larger than the Rydberg induced energy shift.
        All those factors lead to a transfer ratio vastly differing from the full model and demonstrate the importance of the multi-level extension and simultaneous diagonalization of Rydberg-Rydberg and Rabi interaction in the Hamiltonian.

        \section{Full model}
        \label{appsec:full_model}
        Supplementary to the discussion the main paper, this section aims to further elaborate the theoretical description of the full ATI model with its states, the Hamiltonian and the subsequent probing.
        In the following we will use the pair index reference for the pair states introduced in Section\,3 but without explicitly referencing the according subspace if not necessary:
        \begin{equation}
            \begin{split}
            \base[ik]&=\psiq[i]\otimes\,\psis[k]\\
            &=\ket{n_i, l_i, j_i, m_{j, i}}_\mathrm{C}\otimes\ket{n_k, l_k, j_k, m_{j, k}, I_k, m_{I, k}}_\mathrm{S}\\
            &=\ket{\psi_i, \psi_k}\,.
            \end{split}
            \label{eq:prod_basis_fss_idx}
        \end{equation}

        \subsection{Hamiltonian and diagonalization}
        Following the ordering already used in the main text, we start with the Rydberg-Rydberg interaction Hamiltonian $\Hryd$ and only consider dipolar interactions.
        In explicit form it is given by
        \begin{equation}
            \Hryd=\sum_{\ket{\mathrm{r}_i}, \ket{\mathrm{r}_m}}\sum_{\ket{\mathrm{r}_k^*}, \ket{\mathrm{r}_n^*}}\ket{\mathrm{r}_i, \mathrm{r}_k^*}\braket{\mathrm{r}_i|\dipoleop_\mathrm{C}|\mathrm{r}_m}\braket{\mathrm{r}_k^*|\dipoleop_\mathrm{S}|\mathrm{r}_n}\bra{\mathrm{r}_m, \mathrm{r}_n^*}\,,
            \label{eq:hryd_explicit}
        \end{equation}
        where $\dipoleop_\mathrm{C;S}$ is the dipole operator in the respective space.
        As this interaction couples pair state components of both, control and sample atom, it is unique in the sense that it is the only part of the model that acts on both subspaces.
        Additionally, $\Hryd$ acts only on pair states where both sub states are of Rydberg character, leaving every state with $\inter$ character untouched.
        For the extension of the subspace $\psiq$ we neglect the additional quantum numbers $I$ and $m_I$ of the corresponding $\psis$ states as they are irrelevant for the Rydberg-Rydberg interaction.
        Equally, the states of $\left\{\ryd\right\}$ are included in $\left\{\rydaux\right\}$ and extended to the basis of $\psis$.

        The second contribution to the Hamiltonian is the pairwise Rabi coupling of states in $\left\{\inter\right\}$ and $\left\{\rydaux\right\}$ by the coupling field. 
        It factorizes as only states in the sample atom subspace are affected and can thus be written as
        \begin{equation}
            \Hrabi[]=\ident[\mathrm{C}]\otimes\Hrabi[, \mathrm{S}]
            \label{eq:hrabi_sep}
        \end{equation}
        with
        \begin{equation}
            \Hrabi[, \mathrm{S}]=-\sum_{\ket{\mathrm{r}_i}}\sum_{\ket{\mathrm{r}_k^*}}\sum_{\ket{\mathrm{e}_n}}\ket{\mathrm{r}_i, \mathrm{e}_n}\braket{\mathrm{e}_n|\dipoleop\mathbf{\hat{E}}|\mathrm{r}_k^*}\bra{\mathrm{r}_i, \mathrm{r}_k^*} + h.c.\,,
            \label{eq:hrabi_s}
        \end{equation}
        $\dipoleop$ being the dipole operator and $\mathbf{\hat{E}}$ the directional amplitude of the coupling field.
        Pair states with pure Rydberg character are not subjected to this interaction, as are two pair states where the sample atom is in an $\left\{\inter\right\}$ state in both cases.

\begin{figure}[ht]
            \captionsetup[subfigure]{oneside,margin={0.5cm,0cm}}
            \vspace*{10pt}
            \centering
            \begin{subfigure}[t]{0.225\textwidth}
                \caption{}
                \centering
                \includegraphics[width=\textwidth]{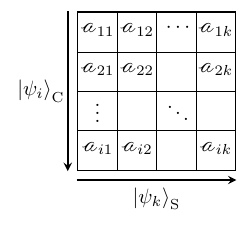}
                \label{sub_fig:eigenstate_matrix_single}
            \end{subfigure}
            \begin{subfigure}[t]{0.225\textwidth}
                \caption{}
                \centering
                \includegraphics[width=\textwidth]{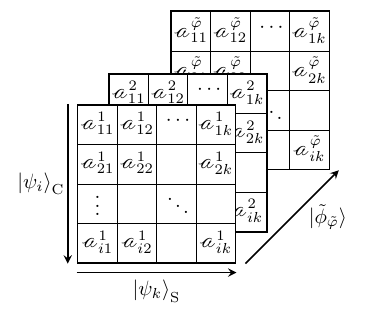}
                \label{sub_fig:eigenstate_matrix_all}
            \end{subfigure}
            \vspace*{-10pt}
            \caption{Base state admixtures in the ATI eigenstates in matrix representation.
            (a) For a single eigenstate $\egbasefs[]$, the size of the two dimensional coefficient matrix is given by the dimensions of the two Hilbert subspaces $\mathcal{H}_\mathrm{C}$ and $\mathcal{H}_\mathrm{S}$.
            (b) Including all eigenstates, the matrix is extended in the third dimension, where the size is given by the dimensionality of the complete Hilbert space $\mathcal{H}=\mathcal{H}_\mathrm{C}\otimes\mathcal{H}_\mathrm{S}$ with $\mathrm{dim}\left(\mathcal{H}\right)=n_\mathrm{c}\cdot n_\mathrm{S}$.
            }
            \label{fig:eigenstate_matrix}
        \end{figure}

        Lastly, the hyperfine interaction for non-Rydberg states needs to be included in the sample subspace.
        It is given by
        \begin{equation}
            \begin{split}
                \Hhfs[]&=\ident[\mathrm{C}]\otimes\Hhfs[, \mathrm{S}]\\
                &=\sum_{\ket{\mathrm{r}_i}}\sum_{\ket{\mathrm{e}_k}, \ket{\mathrm{e}_n}}A_\mathrm{HFS}\ket{\mathrm{r}_i, \mathrm{e}_k}\braket{\mathrm{e}_k|\mathbf{\hat{I}}\cdot\mathbf{\hat{J}}|\mathrm{e}_n}\bra{\mathrm{r}_i, \mathrm{e}_n}\,,
            \end{split}
            \label{eq:hhfs}
        \end{equation}
        where $A_\mathrm{HFS}$ is the hyperfine structure constant of the $\left\{\inter\right\}$ manifold.
        As described in the main text, the ATI Hamiltonian is given by the sum of those three components and transferred to the frame rotating with the coupling field frequency $\omega_c$ before the diagonalization.
        A rotating wave approximation is applied, keeping only the low frequency components of the Hamiltonian.
        Diagonalization of $\Hati$ yields the eigenstates as superpositions of the base states
        \begin{equation}
            \egbasefs[\tilde{\varphi}]=\sum_{i,k}\mathscr{a}_{ik}^{\left(\tilde{\varphi}\right)}\ket{\psi_i, \psi_k}
            \label{eq:eigenstate}
        \end{equation} 
        where the admixture of each base state to the new eigenstate is given by the prefactor $\mathscr{a}_{ik}^{\left(\tilde{\varphi}\right)}=\braket{\tilde{\phi}_{\tilde{\varphi}}|\Psi_{i, k}}$ fulfilling both 
        \begin{equation}
            \sum_{i,k}\left|\mathscr{a}_{ik}^{\left(\tilde{\varphi}\right)}\right|^2=\sum_{\tilde{\varphi}}\left|\mathscr{a}_{ik}^{\left(\tilde{\varphi}\right)}\right|^2=1\,.
            \label{eq:admixutre_normalization}
        \end{equation}
        Using this tuple indexing of the base states, the admixtures for a given eigenstate $\egbasefs[]$ can be written in matrix representation, where the base vectors $\psiq[i]$ ($\psis[k]$) form the rows (columns) of the $n_\mathrm{c}\times n_\mathrm{S}$ matrix, depicted in Fig.\,\ref{fig:eigenstate_matrix}, whose size is given by the dimensionality of the two Hilbert subspaces $n_\mathrm{C}=\mathrm{dim}\left(\mathcal{H}_\mathrm{C}\right), n_\mathrm{S}=\mathrm{dim}\left(\mathcal{H}_\mathrm{S}\right)$.

        Extending the matrix for all eigenstates, we end up with a three dimensional matrix of shape $n_\mathrm{c}\times n_\mathrm{S} \times (n_\mathrm{c}\cdot n_\mathrm{S})$, which in the following is called coefficient matrix with symbol $\mathcal{A}$, where $\mathcal{A}^{\left(\tilde{\varphi}\right)}$ denotes the two-dimensional sub-matrix for a given eigenstate $\egbasefs[\tilde{\varphi}]$ associated with its index $\tilde{\varphi}$.
        The benefit of putting up this kind of coefficient matrix will become apparent in the following when we evaluate the probing of the ATI system.

        \subsection{Probing the system}

        In a first step, we aim to change the base representation of our system, since the ground states $\left\{\g, \gaux\right\}$ are low lying states for which a description in the hyperfine basis is preferable, the Hamiltonian on the other hand was formulated in a fine structure basis.
        Due to the orthonormality of the two bases, the transformation is orthogonal.
        This implies that the eigenenergies are preserved under a basis transformation and the eigenstates in the hyperfine base are simply given by the transformed eigenstates of the fine structure basis.
        The transformation is given by the well-known Clebsch-Gordan decomposition of the fine structure states
        \begin{equation}
            \ket{j, m_j, I, m_I}=\sum_{F, m_F}\ket{F, m_F}\braket{F, m_F|j, m_j, I, m_I}\,.
            \label{eq:cg_decomp}
        \end{equation}
        Since we only probe the $\left\{\inter\right\}$ state manifold, all basis states $\psiq[i]$ are unaffected, making the transformation $T_\mathrm{HFS}^\mathrm{FS}$ separable:
        \begin{equation}
            T_\mathrm{HFS}^\mathrm{FS}=\ident[\mathrm{C}]\otimes T_\mathrm{HFS, S}^\mathrm{FS}\,,
            \label{eq:transformation}
        \end{equation}
        with
        \begin{equation}
            \begin{split}
            T_\mathrm{HFS, S}^\mathrm{FS}=&\sum_{\ket{\mathrm{e}_k^\mathrm{HFS}}}\sum_{\ket{\mathrm{e}_n}}\ket{\mathrm{e}_k^\mathrm{HFS}}\braket{F_k, m_{F, k}|j_n, m_{j, n}, I_n, m_{I, n}}\bra{\mathrm{e}_n}\\
            &\hspace{5em} + \sum_{\ket{\mathrm{r}_k^*}}\sum_{\ket{\mathrm{r}_n^*}}\ket{\mathrm{r}_k^*}\bra{\mathrm{r}_n^*}\,.
            \end{split}
            \label{eq:transformation_explicit}
        \end{equation}
        Making use of the coefficient matrix, changing the base is now a simple matrix multiplication
        \begin{equation}
            \left(\mathscr{A}^{\left(\varphi\right)}\right)^T=T_\mathrm{HFS}^\mathrm{FS}\left(\mathscr{A}^{\left(\tilde{\varphi}\right)}\right)^T\,,
            \label{eq:transformation_product}
        \end{equation}
        or using the orthogonality of the transformation matrix
        \begin{equation}
            \mathscr{A}^{\left(\varphi\right)}=\left(T_\mathrm{HFS}^\mathrm{FS}\left(\mathscr{A}^{\left(\tilde{\varphi}\right)}\right)^T\right)^T=\mathscr{A}^{\left(\tilde{\varphi}\right)}\left(T_\mathrm{HFS}^\mathrm{FS}\right)^{-1}\,,
            \label{eq:transformation_ortho}
        \end{equation}
        where the admixture of the hyperfine structure base states to the transformed eigenstates $\egbase[\varphi]$ is again given by the coefficients of $\mathscr{A}^{\left(\varphi\right)}$.

        For the probing, we are only interested in the state of the sample atom and need to trace out the state of the control atom.
        To this end, the reduced density matrix with respect to the sample atom subspace $\mathcal{H}_\mathrm{S}$ 
        \begin{equation}
            \begin{split}
            \dmatrix[\mathrm{S}]^{\left(\varphi\right)}&=\mathrm{tr}_\mathrm{C}\left(\dmatrix[\mathrm{CS}]^{\left(\varphi\right)}\right)=\mathrm{tr}_\mathrm{C}\left(\ket{\phi_{\varphi}}\bra{\phi_{\varphi}}\right)\\
            &=\sum_{i}\left(\bra{\mathrm{r}_i}\otimes\ident[\mathrm{S}]\right)\ket{\phi_{\varphi}}\bra{\phi_{\varphi}}\left(\ket{\mathrm{r}_i}\otimes\ident[\mathrm{S}]\right)
            \end{split}
            \label{eq:reduced_dm}
        \end{equation}
        can be calculated by multiplying the coefficient matrix with its transposed
        \begin{equation}
            \dmatrix[\mathrm{S}]^{\left(\varphi\right)}=\left(\mathscr{A}^{\left(\varphi\right)}\right)^T\mathscr{A}^{\left(\varphi\right)}
            \label{eq:reduced_dm_matrix}
        \end{equation}
        and has the size $n_\mathrm{S}\times n_\mathrm{S}$.
        Therefore we get $n_\mathrm{S}$ eigenbasis vectors $\dmes[\vartheta]$ when diagonalizing the reduced density matrix with the same amount of eigenvalues $\theta_{\vartheta}$.
        The eigenvalues fulfill the normalization condition
        \begin{equation}
            \sum_{\vartheta}\left|\theta_{\vartheta}\right|^2=1\,,
            \label{eq:norm_cond_rdm}
        \end{equation}
        since by definition $\mathrm{tr}_{\mathrm{S}}\left(\dmatrix[\mathrm{S}]^2\right)=1$ and diagonalization is a trace preserving operation.
        How to get to the transfer probability $\transfer$ from this point is described in the main text.
        For our study we assume the eigenenergies and admixtures to be constant for $\sep\leq\sep_0=0.2\,\si{\micro\meter}$ and keep the values fixed to those calculated at $\sep=\sep_0$ where the system starts to enter the regime at which the Rydberg-Rydberg interaction leads to rather complex spectra due to avoided crossings and the emergence the increasing effect of higher multipole interactions.

        \section{Internuclear distance probability density}
        \label{appsec:idpd}

        As the Rydberg-Rydberg interaction is depends on the distance between the two involved particles also all quantities in our calculations carry this pair distance dependence.
        In an ensemble the distance between randomly picked pairs of particles is determined by the internuclear distance probability density.
        Thus, the sample transfer probability $\transfer$ to be calculated has to take into account the internuclear distance probability density in the ensemble under study. 
        This density is influenced by external and experimental parameters such as the the trap geometry and the temperature $T$ of the atoms.
        Within this work, we assume the trap to be well approximated by a harmonic potential.
        It can thus be described by the trap frequencies $\left(\omega_x, \omega_y, \omega_z\right)$ in the three spatial directions, which, in an optical tweezer, depend on external parameters such as the wavelength $\lambda$ and power $P_t$ of the trapping laser as well as its waist $w_0$ at the position of the atoms.
        In this case, the atomic density distribution of a thermal ensemble is given by
        \begin{equation}
            n\left(\sep, T\right)=n_0\exp\left({-\left(\frac{x^2}{2\sigma_x^2}+\frac{y^2}{2\sigma_y^2}+\frac{z^2}{2\sigma_z^2}\right)}\right)\,,
            \label{eq:density_distribution}
        \end{equation}
        where the width of the sample $\sigma_i$ in the according direction given by
        \begin{equation}
            \sigma_i=\sqrt{\frac{k_\mathrm{B}T}{m\omega_i^2}}\,.
            \label{eq:width}
        \end{equation}
        From this density, it can be shown that the internuclear distance probability density (short IDPD) can be analytically calculated to be \cite{Tu2002RandomDD}
        \begin{equation}
            \dens\left(\sep, T\right)=\frac{R^2}{2\sqrt{\pi}\sigma_x\sigma_y\sigma_z}\mathrm{exp}\left(-\frac{R^2}{4\left(\sigma_x\sigma_y\sigma_z\right)^{2/3}}\right)\,.
            \label{eq:idpd}
        \end{equation}
        Weighting the transfer probability $\transferr\left(\sep\right)$ by the IDPD, the transfer probability of all atoms in the sample is 
        \begin{equation}
            \transfer=\int_0^{\infty}\transferr\left(\sep\right)\dens\left(\sep, T\right)\mathrm{d}\sep\,.
            \label{eq:transfer_sample_app}
        \end{equation}

        \section{Numerical parameters}
        \label{appsec:nummeric_states}

        The ATI scheme presented in this study is very versatile and can be realized for a large variety of atomic species, internal states and external parameters.
        Naturally, we have to restrict our numerical analysis of the model to a certain set of parameters.
        Here, we motivate the particular choice of parameters we took.

        \subsection{Atomic Species and States}
        Due to the large abundance of tweezer experiments working with $^{87}\mathrm{Rb}$, we settled for this element.
        Within \Rb\, the implementation of  $\left\{\g\right\}$ and $\left\{\gaux\right\}$ as the two hyperfine ground state manifolds $5S_\frac{1}{2}\mathrm{F}=1$ and $5S_\frac{1}{2}\mathrm{F}=2$ is straight forward since the information on the control excitation can be stored in a long-lived state and preserved for later detection.
        The choice of the intermediate state, to which the probe transition couples, i.e. $\inter$, has a lot more variability. 
        We chose the $\left\{\inter\right\}=\intnum[; F, m_f]$ manifold, as it can easily be addressed by a $\lambda_p=420\,\si{\nano\meter}$ laser in a well-established two-photon excitation scheme to the Rydberg state.
        Within this set of states, we probe close to the $F=2$ states which provides a decent dipole matrix elements from the $\gsnum[, m_F]$ states while still having a finite branching ratio $\branch[]=0.5$ to the $\left\{\gaux\right\}$ manifold.
        By the nature of two-photon transitions from an $S$ ground state, Rydberg $S$ and $D$ states are possible candidates for the choice of the $\rydaux$ state. 
        As discussed in the main text, the two Rydberg states $\ryd$ and $\rydaux$ have to differ, as otherwise the possible excitation of the control atom will be rapidly mapped down to the ground state by the coupling field.
        Typically larger principle quantum numbers of the Rydberg states provide larger interaction strength.
        At the same time, however, the optical coupling strength reduces due to the reducing dipole matrix element for the coupling from the intermediate state.
        A favorable situation in this respect is created by low-lying Förster resonances, where the resonant energy transfer between (almost) degenerate Rydberg pairstates substantially increases the interaction strength $\vint$.
        The use of Förster resonances is also possible in systems with differing atomic species of control and sample atoms, as they are not restricted to single atomic species \cite{PhysRevA.92.042710, PhysRevResearch.6.013293, PhysRevResearch.2.033474}.
        To demonstrate the capabilities of the imaging scheme at relatively low principle quantum numbers, where the influence of external fields is minimized, we  decide to use the state pair $\rydnum\otimes\rydnumaux[]$ which is close to the resonance
        \begin{equation*}
            \rydnum\otimes\rydnumaux[]\leftrightarrow\rydfo[]\otimes\rydfo[].
        \end{equation*}
        Using higher lying resonances such as
        \begin{equation*}
            \ket{67S_{1/2}}\otimes\ket{69S_{1/2}}\leftrightarrow\ket{67P_{1/2}}\otimes\ket{68P_{3/2}}
        \end{equation*}
        with an even reduced Förster defect $\Delta_\mathrm{F}=2\pi\times 2.83\,\si{\mega\hertz}$ and increased $C_3$ and $C_6$ interaction coefficients, we assume even better contrasts.

        \subsection{Coupling Transition}
        While the Rydberg-Rydberg interaction and the hyperfine interaction are fixed quantities for a given choice of atomic species and states, the parameters of the coupling field can be varied externally.
        The coupling strength is described by the Rabi frequency $\Oc=\frac{1}{\hbar}\braket{\mathrm{r}^*|\dipoleop\hat{\mathbf{E}}|\mathrm{e}}$ which is set by the polarization $q_c$ and the the electric field $\hat{\mathbf{E}}$ of the coupling light.
        For a fixed polarization and light field strength, the coupling is still not constant as a function of $R$, as the overlap of the different $\intnum[; F=2, m_F]$ states with the two $\rydnumaux[, m_j=\pm\sfrac{1}{2}]$ states varies due to the changing $m_j$ composition of the former for varying $m_F$.
        In order to still quantify the coupling field by a single quantity, we choose the weakest of the non-zero couplings as a reference to label our parameter sets.

        \subsection{Probing Transition}
        For the probing of the ATI system, the properties of the probe field as well as the properties of the sample influence the outcome.
        In our analysis, we keep the sample parameters fixed and concentrate on the more easily variable parameters of the probe field.
        The duration of the probe pulse is set to $15\,\si{\micro\second}$ which corresponds to half the lifetime $\tau_\mathrm{Ryd}^{\mathrm{bb}}=32.27\,\si{\micro\second}$ of the $\rydnum[]$  state of the control atom in a black-body environment of $T_\mathrm{env}=300\si{\kelvin}$.
        To further restrict the number of varied parameters, we set the probe polarization to $q_p=+1$ and do not vary the probing strength.
        Additionally, we keep the duration $t_p$ and strength $\op$ fixed because their influence on $\transfer$ is rather straight forward.
        An increase/decrease in either of them increases/decreases the exponent in Eq.\,(22) (main text) and thereby the transfer probability. 
        Whilst a variation in $t_p$ leads to a linear variation of the exponent, a change in the coupling strength leads to a variation approximately proportional to $\sfrac{\op^2}{\left(1+\op^2\right)}$ as it changes the Rabi coupling strength in Eq.\,(18) (main text) which shows the mentioned scaling in Eq.\,(17) (main text).
        The detuning $\Dp$ on the other hand has a non-trivial influence, as it leads to an more (in)efficient coupling to different eigenstates in the eigenspectra depicted in Fig.\,5, hence why we focus on this variation in the frame of this proposal.

        The probing strength $\op$ needs to fulfill certain two criteria.
        It needs to be small compared to the other interactions such as $\vint, \Oc$ and $V_\mathrm{HFS}$ to be able to treat it as a perturbation, i.e. separately from the diagonalization of the system, and the probing transition needs to not be saturated, as otherwise Eq.\,(20) (main text) breaks down.
        In the lights of these conditions, we choose $\op=2\pi\times 1\,\si{\mega\hertz}$ in order to stay quantitative, ensuring both conditions to be satisfied.

        \section{Fidelity}
        \label{appsec:fidelity}

        To calculate the detection fidelity of the ATI scheme, we numerically model the photon scattering of the probe transition on a the sample.
        We aim to obtain histograms for the probability to detect a certain number of probe photons for the two different states of the control atom
        By comparing these two histograms, we extract the fidelity for detecting the state of the control atom.
        In the modelling, we include three random processes: i) the temporally exponentially decaying probability for the control atom to be in the Rydberg state ii) the Poissonian process of transferring an atom into the $\gs$ state under probe light illumination and iii) the Poissonian process of photon scattering of imaging light on the atoms previously transferred into the $\gs$ state.

        To calculate the fidelities, we use the results presented in Fig.\,6 from which we deduce an effective scattering rate 
        \begin{equation}
            R^\mathrm{eff}=\frac{\ln\left(1-\transfer_\mathrm{ml}\right)}{t_p\,\ln\left(\branch[]\right)}
            \label{eq:scat_eff}
        \end{equation}
        that averages over the different scattering channels in the multilevel system.
        $\transfer_\mathrm{ml}$ denotes the quantity to be calculated in Section\,11.
        We determine this effective scattering rate for both cases, the excited control atom and the control atom in the ground state.
        Using this effective rate, the transfer for different times $t_0$ and decay times $t_d\leq t_0$ of the control excitation is modeled similar to the method described in the main text
        \begin{equation}
            \transfer=1-\left(1-\branch[]\right)^{\Nph}
            \label{eq:transfer_app}
        \end{equation}
        but with the number of scattered photons $\Nph$ given by
        \begin{equation}
            \Nph\left(t_0, t_d\right)=\begin{cases}
                R^\mathrm{eff}\cdot t_d + R_\mathrm{noRyd}^\mathrm{eff}\cdot\left(t_0-t_d\right), & \text{if } \ket{\psi}_\mathrm{C}\in\left\{\rydaux\right\} \\
                R_\mathrm{noRyd}^\mathrm{eff}\cdot t_0,  & \text{else}
            \end{cases}\,,
            \label{eq:photon_cases}
        \end{equation}
        where for each probe time $t_0$, we calculate the photon number for a whole range of incrementally changing decay times. 
        For $t_0<10\,\si{\micro\second}$, $5\,\si{\nano\second}$ intervals are used, for longer times we use $25\,\si{\nano\second}$ steps.

        \subsection{State Transfer}
        Having accessed the transfer probabilities $\transfer\left(t_0, t_d\right)$, the probability distribution of transferred atoms starting with a sample of $N_\mathrm{at}$ atoms is described by a Poisson distribution
        \begin{equation}
            \pois\left(\nato, \Nato, t_0, t_d\right)=\pois\left(\nato, \mu=\Nato\cdot\transfer\left(t_0, t_d\right)\right)\,,
            \label{eq:poisson_transfer}
        \end{equation}
        where $\nato$ is the number of transferred atoms.
        As a detection method, we use the statistics of a simple fluorescence imaging protocol for which we assume  to detect on average $\bar{n}_\mathrm{ph}=10$ photons per atom with a probability distribution also modeled by a Poisson distribution, i.e. the probability of detecting $\nph$ photons is in the case of $\nato$ atoms is given by $\pois(\nph, \mu=\nato\cdot\bar{n}_\mathrm{ph})$.
        Combining the two distributions, we get the probability of detecting a given photon number $n_\mathrm{ph}$ by
        \begin{equation}
            \mathcal{N}\left(\nph, \bar{n}_\mathrm{ph}, \Nato, t_0, t_d\right)=\mathcal{N}_\mathrm{at}+\mathcal{N}_\mathrm{bg}\,,
            \label{eq:combined_probability}
        \end{equation}
        where 
        \begin{equation}
            \begin{split}
            \mathcal{N}_\mathrm{at}=\sum_{\nato\geq 1}\pois&\left(\nato, \mu=\Nato\cdot\transfer\left(t_0, t_d\right)\right)\\
            &\cdot\pois(\nph, \mu=\nato\cdot\bar{n}_\mathrm{ph})
            \end{split}
            \label{eq:prob_photons}
        \end{equation}
        accounts for the situation where an atom is transferred and 
        \begin{equation}
            \mathcal{N}_\mathrm{bg}=\pois\left(\nph, \mu=5\right)
            \label{eq:prob_bg}
        \end{equation}
        is used to model the background noise in the detection.

        \subsection{Control Atom Decay}
        If the control atom is in its ground state, we only get a single distribution $\mathcal{N}_\mathrm{noRyd}=\mathcal{N}_\mathrm{noRyd}\left(\nph, \bar{n}_\mathrm{ph}, \Nato, t_0\right)$, as the number of photons scattered in the duration of the probe pulse $\Nph$ is independent of the time of loss of the control excitation $t_d$ since we use the second case in Eq.\,\eqref{eq:photon_cases}.
        With initial excitation, the situation complicates, since the detection probability depends on $t_d$.
        In this case, the different photon distributions are weighted by the probability to lose the excitation at a given time described by the deviation of the survival probability resulting in an exponential continuous random variable
        \begin{equation}
            \mathrm{Exp}\left(t, t_0\right)=\begin{cases}\frac{1}{\tau_\mathrm{Ryd}}\,e^{-t/\tau_\mathrm{Ryd}}, & t<t_0\\
            P\left(t_0\right), & t=t_0\\
            0, & t>t_0
            \end{cases}\,,
            \label{eq:pdf_ryd}
        \end{equation}
        $\tau_\mathrm{Ryd}$ being the lifetime of the excitation and $P\left(t_0\right)$ chosen such that $\int_{0}^{\infty}\mathrm{Exp}\left(t, t_0\right)\mathrm{d}t=1$.
        The photon histogram is therefore given by
        \begin{equation}
            \begin{split}
            \mathcal{N}=&\mathcal{N}\left(\nph, \bar{n}_\mathrm{ph}, \Nato, t_0\right)\\
            &=\int_{t_d}\mathrm{Exp}\left(t=t_d\right)\cdot\mathcal{N}\left(\nph, \bar{n}_\mathrm{ph}, \Nato, t_0, t_d\right)\mathrm{d}t_d
            \end{split}
            \label{eq:weighted_photon_hist}
        \end{equation}
        The two photon number distributions now help us calculate different fidelities.
        The easiest accessible is the fidelity of correctly identifying the state of the control atom independently of the state itself.
        This fidelity is simply given by the overlap integral of the two distributions
        \begin{equation}
            \mathcal{F}\left(\Nato, t_0\right)=\int_{\nph}\sqrt{\mathcal{N}\cdot\mathcal{N}_\mathrm{noRyd}}\,\mathrm{d}\nph\,,
            \label{eq:overlap}
        \end{equation}
        the square root normalizing the overlap as for probabilities $\int P^2\neq 1$ for a normalized probability distribution $P$.
        If one is interested in only correctly identifying one of the states, it is best to use a threshold, splitting the histogram into two parts.
        To find the threshold, the function
        \begin{equation}
            f\left(n_\mathrm{th}\right)=\sum_{\nph\leq n_\mathrm{th}}\mathcal{N}+\sum_{\nph\geq n_\mathrm{th}}\mathcal{N}_\mathrm{noRyd}
            \label{eq:threshold}
        \end{equation}
        is minimized with respect to $n_\mathrm{th}$.
        With this threshold, all probabilities to correctly/falsely detect the initially prepared state can be calculated by summing the corresponding photon distribution in the two intervals whose one boundary is given by $n_\mathrm{th}$.
        In the main text, we take the average fidelity to correctly identify the state as a figure of merit, i.e. 
        \begin{equation}
            \mathcal{F}=\frac{1}{2}\left[\mathrm{P}\left(\ryd, \ryd\right)+\mathrm{P}\left(\gs, \gs\right)\right]\,,
            \label{eq:fidelity_avg}
        \end{equation}
        where $\mathrm{P}\left(x, y\right)$ is the probability of preparing $x$ and detecting $y$.
        Exemplarily, the different values for a probe time of $t_0=15\,\si{\micro\second}$, a Rabi coupling of $\Oc=2\pi\times 31\,\si{\mega\hertz}$ and three different atom numbers are given in Tab.\,\ref{tab:fidelities}.
        \begin{table}
            \caption{Statistics for the state detection with fidelities in different scenarios for a probe time of $t_0=15\,\si{\micro\second}$ and coupling $\Oc=2\pi\times 31\,\si{\mega\hertz}$.}
            \begin{center}
                \makebox[0.48\textwidth]{
                \begin{tabular}{| l || c | c || c | c || c | c |}
                    \hline
                    & \multicolumn{2}{c ||}{} & \multicolumn{2}{c ||}{} & \multicolumn{2}{c |}{} \\[-0.75em]
                    Preparation & \multicolumn{2}{c ||}{$\ryd$} & \multicolumn{2}{c ||}{$\gs$} & \multicolumn{2}{c |}{either} \\
                    & \multicolumn{2}{c ||}{} & \multicolumn{2}{c ||}{} & \multicolumn{2}{c |}{} \\[-0.75em]
                    \hline
                    & & & & & & \\[-0.75em]
                    Detection & $\ryd$ & $\gs$ & $\ryd$ & $\gs$ & overlap & Eq.\,\eqref{eq:fidelity_avg} \\
                    & & & & & & \\[-0.75em]
                    \hline\hline
                    & & & & & & \\[-1.0em]
                    $\Nato=1$ & 0.356 & 0.644 & 0.138 & 0.862 & 0.066 & 0.609 \\
                    & & & & & & \\[-0.75em]
                    $\Nato=10$ & 0.859 & 0.141 & 0.030 & 0.970 & 0.585 & 0.914 \\
                    & & & & & & \\[-0.75em]
                    $\Nato=100$ & 0.972 & 0.028 & 0.005 & 0.995 & 0.852 & 0.984 \\
                    \hline
                \end{tabular}}
            \end{center}
            \label{tab:fidelities}
        \end{table}

\end{document}